\documentclass[10pt,journal,compsoc]{IEEEtran}
\ifCLASSOPTIONcompsoc
\usepackage[nocompress]{cite}
\else
\usepackage{cite}
\fi
\ifCLASSINFOpdf
\else
\fi
\usepackage{soul}
\usepackage{color}
\usepackage{graphicx}
\usepackage{caption}
\usepackage{subcaption}
\usepackage{tikz}
\usepackage{multirow}
\usepackage{ragged2e}
\usepackage[ruled]{algorithm2e} % For algorithms
\definecolor{amber}{rgb}{1.0, 0.75, 0.0}
\newcommand*\circled[1]{\tikz[baseline=(char.base)]{
\node[shape=circle,draw,inner sep=1.5pt] (char) {\textcolor{black}{#1}};}}

\begin{document}
\title{RC-RNN: \underline{R}econfigurable \underline{C}ache Architecture for Storage Systems Using \underline{R}ecurrent \underline{N}eural \underline{N}etworks}
% received ..." text while in non-compsoc journals this is reversed. Sigh.

\author{Shahriar Ebrahimi,
Reza Salkhordeh,
Seyed Ali Osia,
Ali Taheri,
Hamid R. Rabiee,
and Hossen Asadi
\IEEEcompsocitemizethanks{\IEEEcompsocthanksitem All of the authors are associated with the Department of Computer Engineering at Sharif University of Technology, Tehran, Iran.\protect\\
E-mails: shebrahimi@ce.sharif.edu,~salkhordeh@ce.sharif.edu,
~osia@ce.sharif.edu,~taheri@ce.sharif.edu,~
rabiee@sharif.edu,~asadi@sharif.edu.}% <-this % stops an unwanted space
%\thanks{Manuscript received April 19, 2018; revised August 26, 2019.}
}

% ^------------^------------^----Do not want these spaces!

\markboth{Transactions on Emerging Topics in Computing (TETC)}
{Ebrahimi \MakeLowercase{\textit{et al.}}: RC-RNN: Reconfigurable Cache Architecture for Storage Systems Using Recurrent Neural Networks}
% *** name in the headers of peer review papers. ***



\IEEEtitleabstractindextext{%
\justify
\begin{abstract}
\emph{Solid-State Drives} (SSDs) have significant performance advantages over traditional \emph{Hard Disk Drives} (HDDs) such as lower latency and higher throughput. Significantly higher price per capacity and limited lifetime, however, prevents designers to completely substitute HDDs by SSDs in enterprise storage systems. SSD-based caching has recently been suggested for storage systems to benefit from higher performance of SSDs while minimizing the overall cost. While conventional caching algorithms such as \emph{Least Recently Used} (LRU) provide high hit ratio in processors, due to the highly random behavior of \emph{Input/Output} (I/O) workloads,
they
hardly provide the required performance level for storage systems. In addition to poor performance, inefficient algorithms also shorten SSD lifetime with unnecessary cache replacements. Such shortcomings
motivate
us to benefit from more complex non-linear algorithms to achieve higher cache performance and extend SSD lifetime.

In this paper, we propose \emph{RC-RNN}, the first reconfigurable SSD-based cache architecture
for storage systems
that utilizes machine learning
to identify performance-critical data pages for
I/O
caching.
The proposed architecture uses \emph{Recurrent Neural Networks} (RNN) to characterize ongoing workloads and optimize itself towards higher cache performance while
improving
SSD lifetime. \emph{RC-RNN} attempts to learn characteristics of the running workload to predict its behavior and then uses the collected information to identify performance-critical data pages to fetch into the cache.
We implement the proposed architecture on a physical server equipped with a \emph{Core-i7} CPU, 256GB SSD, and a 2TB HDD running Linux \emph{kernel 4.4.0.} Experimental results
show that \emph{RC-RNN} characterizes workloads with an accuracy up to 94.6\% for SNIA I/O workloads. \emph{RC-RNN} can perform similarly to the optimal cache algorithm by an accuracy of 95\% on average, and outperforms previous SSD caching architectures by providing up to 7x higher hit ratio and decreasing cache replacements by up to 2x.
\end{abstract}

}

\maketitle

\IEEEdisplaynontitleabstractindextext

% \fi
\IEEEpeerreviewmaketitle

\section{Introduction}\label{sec:intro}

Storage subsystems have significant impact on the overall performance of enterprise \emph{Input/Output} (I/O) intensive applications. The major performance bottleneck of storage subsystem is mechanical storage devices such as \emph{Hard Disk Drives} (HDDs), which suffer from limited response time. With emergence of \emph{Flash-based Solid-State Drives} (SSDs) that have no mechanical components, the performance of storage subsystem can be significantly improved. SSDs, however, have several drawbacks such as reliability concerns~\cite{ahmadian2, kishani} and one order of magnitude higher price per capacity compared to HDDs~\cite{reca}.

To alleviate the shortcomings of SSDs while exploiting their benefits, employing SSD as an I/O cache for HDD-based storage subsystems has been studied in the previous studies~\cite{arc, larc, marc, azor, hystor, srac, dura, reca}. In SSD-based I/O caching architectures, frequently accessed requests are buffered in the SSD to provide fast response time while other requests are supplied by HDDs.
Several application domains such as \emph{Database Management Systems} (DBMS), mail servers, \emph{OnLine Transaction Processing} (OLTP), and \emph{High Performance Computing} (HPC) can benefit from SSD-based I/O caching~\cite{reca, larc, azor}. Each aforementioned domain, however, has a distinct workload characteristic and therefore, requires a specific caching policy to fully exploit the benefits of SSD.

\begin{figure*}[t]
\begin{subfigure}[t]{0.49\textwidth}
\centering
\includegraphics[width=\linewidth]{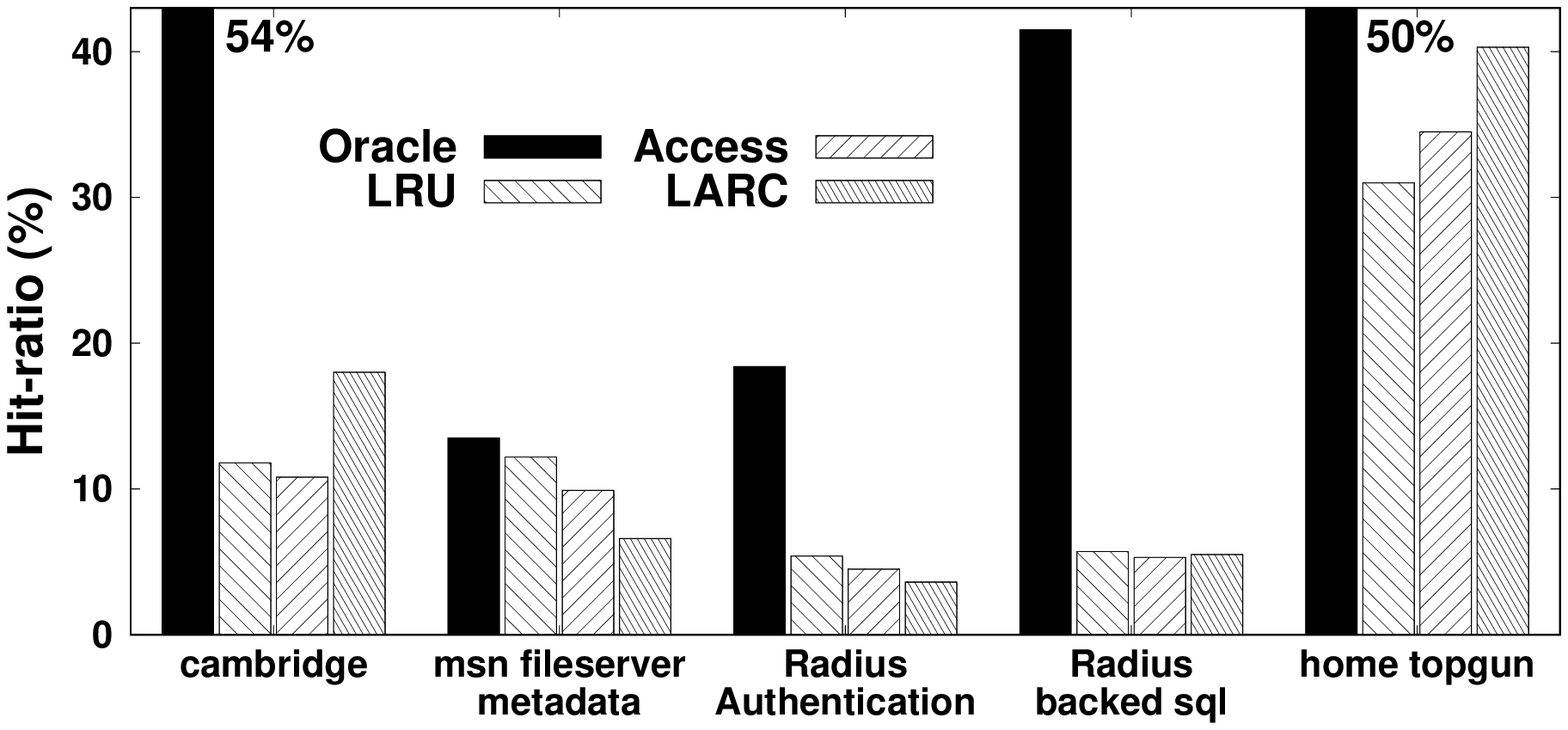}
\caption{Hit Ratio} \label{fig:intro_motiv_a}
\end{subfigure}
\begin{subfigure}[t]{0.49\textwidth}
\centering
\includegraphics[width=\linewidth]{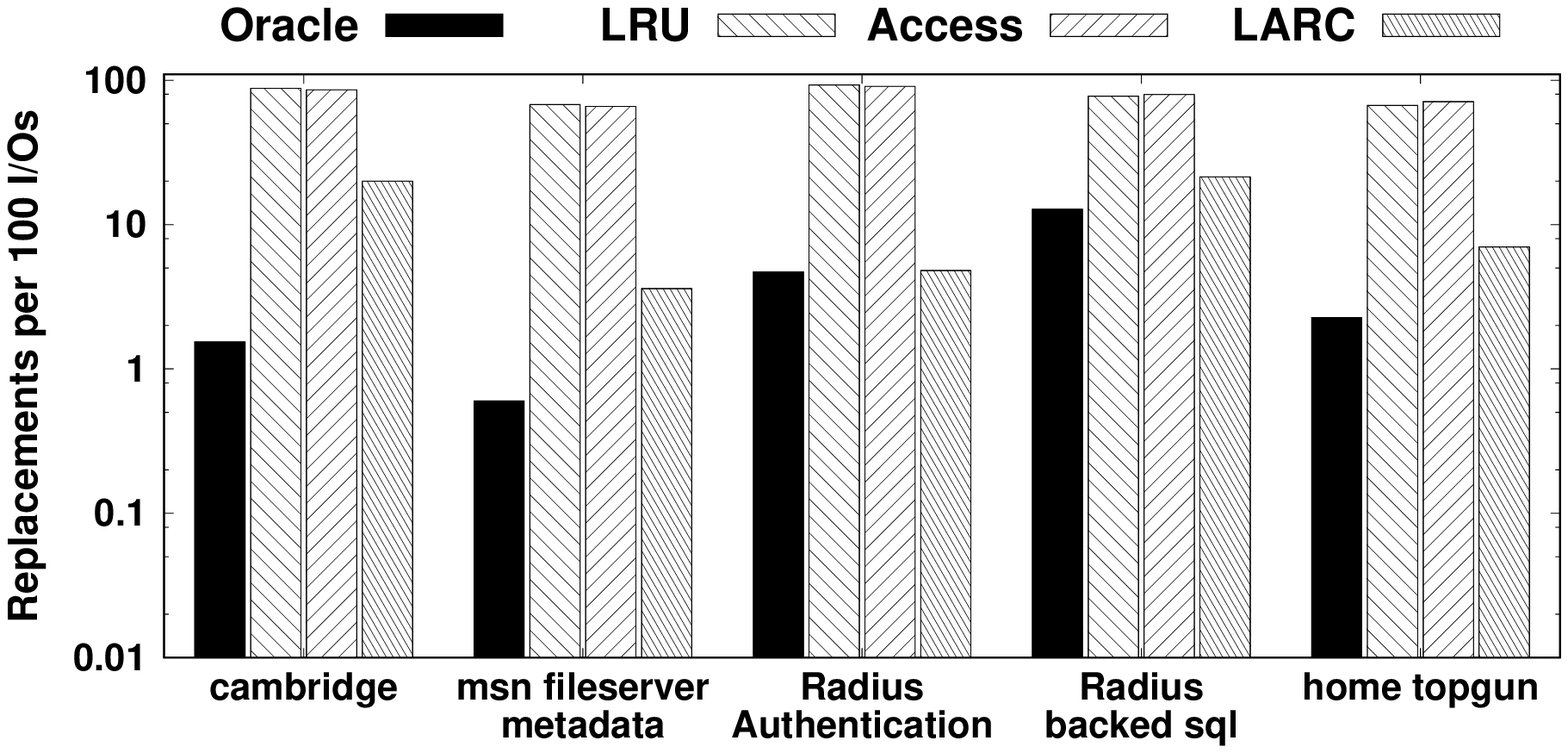}
\caption{Cache Replacements} \label{fig:intro_motiv_b}
\end{subfigure}
\caption{Oracle Compared to Conventional Algorithms} \label{fig:intro_motiv_all}
\vspace{-4mm}
\end{figure*}

Flash cells in SSDs need to be erased before writing.
Such cells can \emph{only} endure a limited number of erases and hence, have a limited lifetime.
Each cache replacement requires writing a new data page to the SSD and therefore, the number of cache replacements directly affects the SSD lifetime. 
%Each missed read results in copying the data page to the cache, which imposes a write on the SSD.
%Improving read hit ratio reduces the number of data pages copied to the cache and therefore, reduces the number of writes into the SSD. 
Moreover, Due to the limited and expensive capacity of SSDs compared to HDDs, the cost of buffering I/O requests is very high. Therefore, accurately identifying the performance-critical data pages is crucial for a cost-efficient SSD-based caching architecture. Hence, for efficiently managing the
I/O cache to improve performance while extending SSD lifetime, caching architectures should: (1) have accurate knowledge about the behavior of the current workload, and (2) be able to reconfigure themselves in case of workload changes. Such reconfiguration can be conducted by providing feedback, based on changes in hit the ratio \cite{arc, larc, marc}. Employing static flags for requests to distinguish metadata from data requests has also been suggested in the previous studies \cite{azor, hystor}. Moreover, in architectures such as~\cite{8550676, reca}, cache priorities are updated periodically based on the workload type. 
In addition, several previous studies aim to reduce the number of writes in SSD to increase SSD lifetime even at the cost of performance degradation \cite{dura, larc}. %\textcolor{red}{In all caching scenarios, there are two sources of writes on the cache: 1)~the write requests from workload itself, and 2)~the read misses in cache forcing an immediate fetch of the missed page from slower device to cache. To increase SSD lifetime, some studies, have tried to force policies, such as \emph{read-only} for certain workloads~\cite{reca}. To have a fair comparison and in order to push caching policies to their limits, in this paper, we have configured all caching architectures in write-back mode. Thus, all of the workload-based write requests are equal in all caching architectures and the number of SSD writes caused by read misses determines the SSD lifetime.}

A major limitation of existing SSD-based caching architectures such as \cite{larc,azor,arc,marc} is that they are adapted from conventional caching algorithms employed in either main memory or processors cache levels. Such algorithms depend on high temporal and spatial locality while I/O workloads mostly exhibit unpredictable behavior with no linear locality \cite{char_tarihi1}, which degrades cache performance and SSD lifetime by issuing inefficient cache replacements. To show the inefficiency of previous studies in I/O workloads, we run experiments to compare hit ratio of \emph{Least Recently Used} (LRU), \emph{Access Frequency} \cite{arc}, and \emph{LARC} \cite{larc} with the \emph{Optimal} Cache replacement policy~\cite{belady} (Oracle/Belady) under four different I/O traces from \emph{Storage Networking Industry Association} (SNIA) \cite{snia}. \color{black}{Figure}~\ref{fig:intro_motiv_a}{ and Figure}~\ref{fig:intro_motiv_b} {show the hit ratio and the number of cache replacements conducted for conventional algorithms as opposed to the \emph{Oracle} (i.e., optimal) algorithm, respectively. Although conventional algorithms have high hit ratio in workloads with high locality such as \emph{mail\_index}, they still impose significant number of unnecessary cache replacements, which decreases SSD lifetime. Moreover, under workloads such as \emph{Radius\_Backed\_SQL}, conventional algorithms fail to provide high hit ratio due to the low linear locality in I/O requests.~\cite{reca} }\color{black}
These experiments reveal that conventional algorithms are \emph{not} efficient for SSD-based I/O caching and impose high number of unnecessary cache replacements in many scenarios.

In addition, several existing architectures: a) are optimized towards a selection of workloads \cite{hystor, larc, marc, azor} \emph{or} b) modify standard filesystems to provide information about request types \cite{azor, hystor}, which makes them heavily dependent on a particular filesystem. Unlike CPU level caches, the average response time of I/O requests is several milliseconds, which provides sufficient \emph{thinking time} to execute complex computations to efficiently identify and predict workload behavior. Such predictions can help caching architectures to reduce the performance and endurance gap between optimal and conventional algorithms. To the best of our knowledge, \emph{none} of the previous studies have utilized machine learning methods to achieve higher cache performance in storage systems
while preserving the cache lifetime intact.

In this paper, we propose the first reconfigurable SSD-based cache architecture, called RC-RNN, which tries to improve both performance and endurance of SSD-based caches by employing \emph{Machine Learning} to: \textbf{a)} identify workload type, \textbf{b)} decide which data pages should be buffered, and \textbf{c)} decide which data pages are no longer beneficial to be buffered.
RC-RNN proposes several request characteristics, which are used by the \emph{Machine Learning} method to accurately identify the running workload.
For each workload type, a \emph{Machine Learning} model is constructed, which performs very similar to the \emph{Optimal} cache policy by deciding to ignore or buffer miss accesses and evicting the cold data pages from the cache.
The proposed cost function employed to construct the \emph{Machine Learning} model considers several workload and storage device characteristics in order to accurately estimate the cost and benefit of buffering data pages.

We utilize \emph{Recurrent Neural Network} (RNN) in the proposed architecture as one of the most powerful machine learning methods, which is proven to be accurate in several application domains such as text analysis~\cite{ml4} and speech recognition~\cite{ml2}.
Finding patterns in a trace of the I/O requests has similarities to both of the mentioned application domains, which encouraged us to select RNN as the machine learning method.
RC-RNN consists of offline and online phases.
\textcolor{black}{Any time-, CPU-, and memory-consuming operations are placed in the \emph{offline} phase and are done only \emph{once}.
In the online phase, RC-RNN monitors the running workload and decides which data pages should be buffered
or evicted based on the constructed models. To improve SSD lifetime, RC-RNN does not copy all
missed data pages to the cache. The data pages with low probability of access in the future are
ignored and responded directly by HDD to reduce the number of writes in SSDs.}

%In this paper, we propose the first reconfigurable SSD-based cache architecture called \emph{RC-RNN}, that uses \emph{Machine Learning} to recognize the implicit patterns of the workload data and reconfigure itself based on them. In order to consider the sequential nature of workload data, we utilize Recurrent Neural Network (RNN)~\cite{ml1} as one of the most powerful machine learning methods which achieves state of the art results in many different applications based on sequential data e.g. text analysis~\cite{ml4}, speech recognition \cite{ml2} and etc. Employing RNNs provides us with (1) characterizing ongoing workload, and (2) dynamically managing I/O requests based on the workload type. We used more than 17 workloads from various applications and for each workload type, a deep RNN model learns the optimal eviction policy to manage the I/O cache under examined workload.

%By employing the proposed RNN model, \emph{RC-RNN} a) decides to whether or not cache a data page, and b) decides when to evict a data page. \emph{RC-RNN} consists of an offline and an online phases. All time-consuming operations are done only once in the offline phase such as training RNN models. In the online phase, \emph{RC-RNN} monitors workload and decides which data pages should be cached or evicted based on the constructed models. To improve SSD lifetime, \emph{RC-RNN} does not copy all missed data page to the cache. The data pages with low probability of access in the future are ignored and responded directly by HDD to reduce the number of writes in SSDs.\\

\emph{RC-RNN} is designed to be reconfigurable with negligible reconfiguration cost.
\color{black}
To this end, it monitors I/O requests online and evaluates the workload time periodically offline. Upon detection of any change in the workload type, \emph{RC-RNN} switches the loaded RNN model to match the new workload type. Note that the reconfiguration process is accomplished only by a swap of small RNN models~(around 40 MB) in memory, which also can be loaded simultaneously due to low memory consumption. 
\color{black}
The internal state of cache is also updated to reflect the workload change. The reconfiguration process enables \emph{RC-RNN} to have high accuracy in various workload types unlike previous studies which are optimized toward a few workloads~\cite{azor,arc,larc}.

\textcolor{black}{We evaluate \emph{RC-RNN} by using real SSDs and HDDs
in a server equipped with \emph{Core-i7} CPU running \emph{Linux 4.4.0} on \emph{Ubuntu 16.04} operating system.
RNN models are implemented by using the \emph{Keras} \cite{keras} library and the \emph{online} phase experiments are evaluated \emph{only} by CPU.} We show that the proposed characterization method has up to 94.5\% accuracy in detecting workloads from SNIA I/O traces~\cite{snia}.
Experimental results show that \emph{RC-RNN} can perform similarly to Oracle algorithm with less than 5\% error in both preventing seldom accessed data pages from entering the cache and timely deciding the data pages to evict.
\emph{RC-RNN} improves hit ratio by up to 7x (2x on average) compared to state-of-the-art caching architectures. By filtering seldom accessed data pages, \emph{RC-RNN} reduces the number of writes in SSD by up to 22\%
(1.7\% on average) 
compared to previous studies.

In summary, the \textbf{main contributions} of this work are as follows:
\begin{itemize}
\item We propose the first comprehensive I/O workload characterization method, which considers long dependencies between requests. Unlike previous workload characterization methods, which \emph{only} consider a limited number of past requests, the proposed method can identify several streams of requests originating from a single or multiple applications.
\item We introduce RC-RNN, the first I/O caching architecture capable of employing machine learning methods to decide which data pages should be placed in or evicted from the cache \textcolor{black}{ in a dynamic priority-based caching policy}. RC-RNN constructs a model from an optimal cache policy and employs it in the runtime to decide suitable cache operations for the running workload.
\item To improve the accuracy of the machine learning model employed in RC-RNN, we propose a cost function that considers the characteristics of SSDs as compared to HDDs to estimate the actual cost of caching, by-passing, or evicting a data page.
\item RC-RNN monitors the running workload using the constructed RNN model for workload characterization, where once a change in the workload is detected, the employed RNN model for cache policy is dynamically changed. The performance impact of the reconfiguration process is kept as minimum as possible.
\item \textcolor{black}{We implement RC-RNN in a real system and show that the proposed architecture improves cache hit ratio up to 7x (2x on average) with negligable overhead during \emph{online} phase compared to previous work.}
\end{itemize}

The rest of the paper is organized as follows. In Section \ref{sec:related}, the background and previous studies are presented. Section \ref{sec:motiv} discusses motivations of this paper. In Section \ref{sec:arc}, the proposed architecture is detailed. Experimental results are presented in Section \ref{sec:exp_res}. Finally, Section \ref{sec:con} concludes the paper.
\vspace{-5mm}
\section{Background and Related Works}\label{sec:related}
In this section, we first discuss emergence of new technologies for storage devices and the motivations to employ heterogeneous hybrid architectures. Next, the suggested hybrid architectures in previous studies are detailed. Current I/O workload characterization methods are also discussed in this section. Finally, a summary of machine learning methods, applicable to I/O caching is provided.

\subsection{Hybrid Architectures}
With emergence of SSDs having performance advantages over mechanical HDDs, the average response time of storage devices has been decreased significantly. However, the higher price of SSDs compared to HDDs and limited lifetime prevents data centers to completely replace HDD-based storage subsystems with SSDs. 

Multi-tiered storage subsystems have been suggested by previous studies to exploit the advantages of both mechanical and non-mechanical storage devices \cite{arc, azor, larc, marc, hystor, srac, dura, tiering1, integrating}. There are two main approaches in hybrid storage systems: (1) \emph{Tiering} and (2) \emph{Caching}. In tiering, SSD is placed at the same level as HDD and the overall capacity of the system is equal to sum of the space of both devices \cite{tiering1, integrating}. Contrary to tiering, in caching a copy of data page is moved to SSD to speed up I/O requests. Tiering is more efficient in terms of overall cost and capacity compared to caching, but due to costly data migration between tiers, it has lower performance during abrupt workload changes. Therefore, tiering is more suitable for systems with steady workloads \cite{tiering2, Guerra}. On the other hand, a caching approach achieves higher performance in workloads with sudden changes, which makes them efficient on systems running simultaneous workloads.

In this paper, we focus on caching since it is capable of responding faster to workload changes. Due to the unpredictable and non-linear behavior of I/O workloads, the performance of an SSD-based caching approach highly depends on the caching algorithm. Using an inefficient caching algorithm not only degrades performance, it also shortens SSD lifetime because of the unnecessary cache replacements. To achieve efficient caching management, it is necessary to predict behavior of workload and be able to configure caching architecture based on the access pattern. In addition, in case of workload change, previously configured caching algorithm may not be efficient any more. Hence, it is important to have a reconfigurable architecture, which can adapt to rapid changes in the workload.

%\begin{table} 
%\caption{HDD vs. SSD} 
%\begin{center} 
%\begin{tabular*}{\linewidth}[t]{ l c c } 
%& \textbf{HDD} & \textbf{SSD}\\ 
%\hline
%Model & WD Red Pro & Samsung 850 PRO \\ 
%Sequential r/wr & 120/120 MB/s & 512/450 MB/s\\ 
%Random r/wr & 80/150 IOPS & 35/83 MB/s\\ 
%Price (per GB) & 0.03\$ & 0.50\$ \\ 
%Number of wites & Unlimited & Limited\\ 
%Vibration & Platters & None\\ 
%Energy (watts)& 6 $\sim$ 15 & 2 $\sim$ 5\\ 
%\end{tabular*} 
%\end{center} 
%\label{tab:hdd_ssd} 
%\end{table} 
\vspace{-2mm}
\subsection{Previous Hybrid Architecture} \label{subsec:cch_arc}
There are numerous studies in SSD-based caching for storage systems \cite{arc, azor, larc, marc, hystor, srac, dura, ahmadian1}. Most of these studies suggest simple and linear caching algorithms based on conventional algorithms such as \emph{Least Recently Used} (LRU) or \emph{Access Frequency} that have shown to be practical and efficient in other levels of memory hierarchy such as CPU caches. Although LRU has low overhead, certain workloads can cause cache thrashing~\cite{thrashing, reca} and significantly decrease its efficiency.

Previous studies that use LRU as their baseline caching algorithm tried to overcome this problem with various approaches. ARC \cite{arc} employs two lists to keep track of recency and frequency of accessed I/O blocks. It buffers a percentage of each list based on the hit ratio in the recent accesses and keeps balance between blocks that have been accessed recently and blocks, which have been mostly accessed during the workload. \emph{LARC} \cite{larc} employs a virtual LRU queue as a filter for recently accessed blocks and prevents randomly accessed blocks from entering cache in the first access. If a block is accessed for the second time while in filter, it gets promoted to the main LRU and will be buffered. With this approach, \emph{LARC} prevents cache thrashing and significantly decreases cache replacements compared to LRU. \emph{mARC} \cite{marc} benefits from both previous architectures in different phases of the workload. It uses hit ratio, recency, and frequency as feedback to switch between two methods. Although the mentioned studies have improved SSD caching performance in few workloads, they still have low performance in other workload types. In addition, the number of unnecessary cache replacements is still high in such architectures.

In addition to modifying LRU, using access frequency has also been suggested in the previous studies \cite{azor, hystor}. In \cite{azor}, a caching architecture called Azor is suggested, which is an access frequency cache that assigns metadata blocks a higher priority over data blocks. This architecture will be effective only for workloads with high percentage of metadata accesses. In case of workload change, older metadata blocks, which are not performance critical anymore, will be kept in cache and prevent performance critical pages from entering the cache. \emph{Hystor} \cite{hystor} proposes a hybrid storage architecture consists of both SSD and HDD to improve the overall performance. One of the drawbacks of \emph{hystor} is allocating a fixed section of SSD as a writeback buffer for HDD subsystem, which can be either too large or too small for certain workloads. The size of the write-back cache significantly affects the performance of running workloads.

\subsection{Workload Characterization} \label{subsec:wrk_char}
The performance of caching architectures in storage systems is highly affected by the characteristics of the running workload. Therefore, workload analysis plays a key role in studies aimed at improving performance of I/O subsystems. The behavior of enterprise application workloads is very complex and difficult to characterize because the performance of each request depends on the previous ones. The non-linear behavior of I/O workloads and burstiness of requests~\cite{char_tarihi2, reca} add to the complexity of characterization. 

To understand the I/O workload characteristics, several previous studies have proposed different analytical models for analyzing enterprise I/O traces \cite{char_disk_io,char_ms_storage,char_peta,char_riska,char_web}. \emph{Read/write ratio}, \emph{I/O size}, and \emph{inter-arrival times} are a few of the common parameters employed in previous studies for workload characterization. In \cite{char_disk_io}, spatial locality and outstanding I/Os are considered in characterization of workloads for HDD-based storage systems. In \cite{char_riska}, workloads are divided into three major domains: enterprise, desktop, and consumer electronics. This work shows that burstiness as a temporal locality measure is highly application dependent and significantly affects the overall workload performance. In \cite{char_web}, six different web servers are characterized in detail and two proposed caching strategies for web caches are obtained through data analysis. Although several parameters such as \emph{read/write ratio, I/O size, temporal} and \emph{spatial} localities, and \emph{burstiness} have been examined in previous studies, they still cannot predict the I/O workload behavior with high accuracy. As we show in Section~\ref{subsec:motiv_char}, this is due to the long dependency of requests to previous requests, which cannot be captured by state-of-the-art characterization methods. 

\subsection{Machine Learning - RNN}\label{subsec:ml1}

In recent years, artificial intelligence has been extensively used in a broad range of applications such as multimedia signal processing (e.g., image, video and speech processing), intelligent systems (e.g., autonomous cars and smart homes) and bio-informatics (e.g., DNA analysis)~\cite{deep, ml2,  ml4, ml5}. Such applications need complicated analysis of large volume of data, which is usually achieved by machine learning methods that recognize the complex implicit patterns of the data and use it for future prediction. In general, building a machine learning model has two phases: \emph{training} and \emph{testing}. In the training phase, which is done offline, a large volume of training data is analyzed to solve an optimization problem for learning a model. In the test phase, the trained model is used to predict features of unseen new data. Among machine learning methods, \emph{Deep Learning} has made great progress in recent years and has achieved state-of-the-art results in many problems~\cite{ml2, ml4, ml5}. \emph{Deep Convolutional Neural Networks} (CNNs) and \emph{deep Recurrent Neural Networks} (RNNs) are the most useful architectures in this field. The former is used to analyze image data and the latter is employed for sequential data analysis such as text and time series data~\cite{label}. We are going to use RNNs in this paper.

%In order to design an efficient reconfigurable cache architecture, we need to extract patterns existed in workload data and use it for future prediction. Since workload data are sequential time series data, we have utilized them and build a learning model based on 

\begin{figure}[t]
\centering
\begin{subfigure}[t]{0.76\linewidth}
\centering
\includegraphics[width=\linewidth]{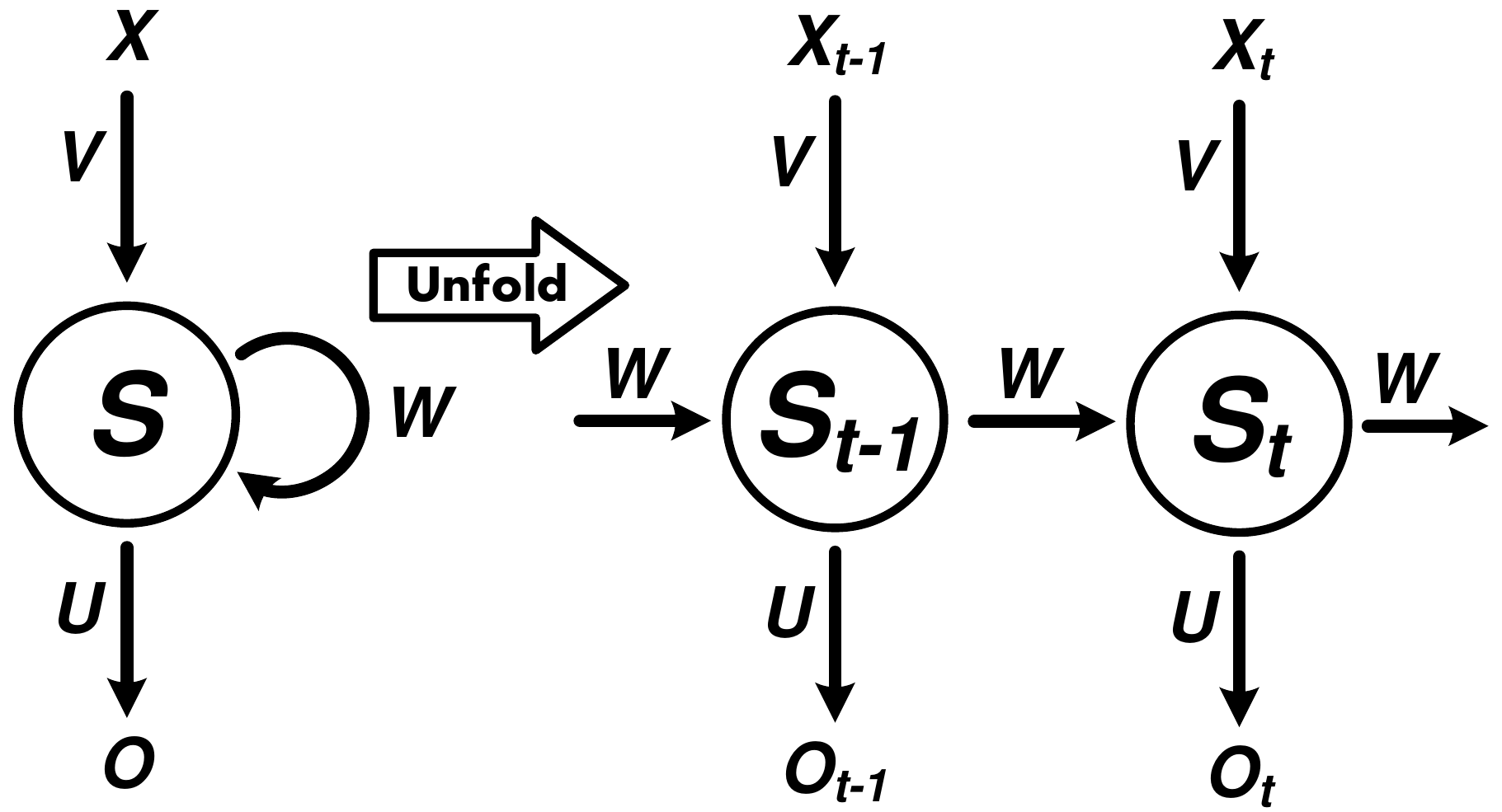}
\caption{Simple RNN} \label{fig:motiv_rnn_fold} 
\end{subfigure}
\quad
\begin{subfigure}[t]{0.19\linewidth}
\centering
\includegraphics[width=\linewidth]{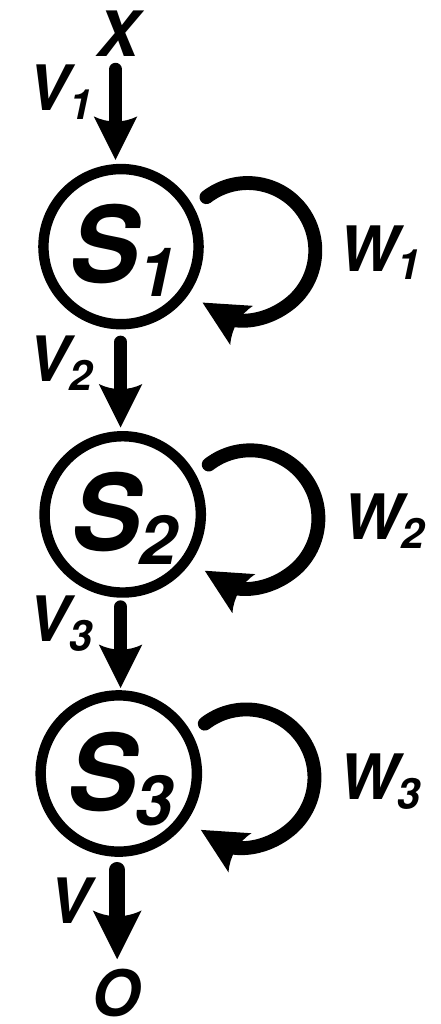}
\caption{Deep RNN} \label{fig:motiv_rnn_deep}
\end{subfigure}
\caption{Different RNN Architectures [15]}\label{fig:rnn_all}
\end{figure}

RNN attempts to model a nonlinear dynamical system with input $x(t)$, inner state $s(t)$, and output $y(t)$. As an example in data storage systems, the input data $x(t)$ can be considered as a request at time $t$ from an I/O workload trace and $y(t)$ indicates the caching mechanism suitable for individual requests. In this scenario, the state $s(t)$ plays the role of the memory and tries to summarize all of the past events ($x(1),...,x(t-1)$). RNN implements the following scenario in a recursive manner: 1) take the new input data, 2) check the previous state, 3) mix them together to build the current state, and finally 4) make the new output by manipulating the current state:
\begin{equation} \label{x_s_y}
s(t) = f(s(t-1),x(t))$$$$
y(t) = g(s(t))
\end{equation}
In order to model the non-linearity of the system, $f$ and $g$ are usually obtained by combining a linear transformation and a simple nonlinear function, e.g., \emph{sigmoid}. Since RNNs, unlike feed-forward neural networks, can use their internal memory to process arbitrary sequences of inputs, they are more suitable for analyzing I/O traces. In RNNs, connections between the nodes form a directed cycle to imitate a dynamic temporal behavior. RNNs can use their internal memory to process arbitrary sequences with input, inner \emph{state}, and output layers. A simple RNN with one hidden layer as the inner state takes the sequential input and updates its current state based on the current input and previous state. The inner state plays the role of the memory and tries to summarize the input data. The output which could be the result of classification or regression, is produced by manipulating the state of the system. Since natural dynamical systems are non-linear, RNN also uses simple nonlinear functions such as \emph{sigmoid}, \emph{tanh}, or \emph{max}, and builds a complex non-linear function by combining them \cite{deep}. A schematic of a simple RNN is shown in Figure \ref{fig:motiv_rnn_fold}. Equation \ref{eq:rnn_base} shows the relation between input ($x$), state ($s$), and output ($u$) of a typical RNN. Where $g$ and $f$ are the non-linear functions (in classification, common choices for these functions are the \emph{rectifier linear unit} and \emph{softmax} function), and $A$, $B$, and $C$ are the weights, which should be learned during the training process.
\begin{equation} \label{eq:rnn_base}
s(t) = f(A.s(t-1) + B.x(t)),\hspace{4mm} u(t) = g(C.s(t))
\end{equation}

For training RNNs, we should first prepare suitable training datasets, which consist of sufficient input sequences with their correct ground truth label. The training is achieved by using an efficient implementation of \emph{Stochastic Gradient Descent} (SGD) algorithm, called \emph{back propagation through time}. We obtain the best estimates for the weights through the training process, but the model should be evaluated by using test datasets.
Different challenges, e.g. vanishing and exploding gradients, might be raised during the training of a simple RNN. These challenges can be resolved by using a special kind of memory unit called \emph{Long Short Term Memory} (LSTM)~\cite{lstm} in the RNN architecture. LSTM defines the inner state with a more complex process, with input, output, and forget gate. The details of this memory unit can be found in \cite{lstm}. Increasing the number of hidden layers and building a deep RNN is an important generalization of RNN, which can handle more complex data patterns at the cost of more time and memory consuming training process. Figure \ref{fig:rnn_all} shows the schematic of a simple RNN (Figure \ref{fig:motiv_rnn_fold}) and a 3-layered deep RNN (Figure \ref{fig:motiv_rnn_deep}), respectively.%which is used in Section \ref{sec:exp_res} in the cache management procedure. 

\color{black}{RNN models are mostly used on one dimensional datasets.
However, RNN can also be employed on multi-dimensional datasets by feeding \emph{d}-dimensional vector (e.g. [addr, size, ...]) at each time step as shown in several previous studies such as}~\cite{andermatt2016multi}.
{RC-RNN uses the same technique to feed the multi-dimensional dataset to the RNN model.
For instance, we use a 100*4 matrix for 100 consecutive time steps to classify the characteristics of the workload based on \emph{four} different dimentions.}
\color{black}

\vspace{-2mm}
\section{Motivation}\label{sec:motiv}
The motivation for this work is three-fold. First, we show the gap between state-of-the-art and optimal (\emph{Oracle}) caching architectures in terms of performance and SSD endurance. Second, we demonstrate the inaccuracy of previous workload characterization methods in several scenarios. Finally, we present several examples of the potential benefits of employing machine learning in both I/O workload characterization and SSD-based I/O cache management.

\vspace{-1mm}
\subsection{Cache Management}\label{subsec:motiv_cch}

To evaluate the efficiency of caching architectures, hit ratio should be compared to the hit ratio of \emph{Oracle} caching architecture, which is proven to have the maximum possible hit ratio due to having the knowledge about all future accesses \cite{belady}. Figure \ref{fig:intro_motiv_a} shows the hit ratio of three different algorithms: 1) \emph{LRU}, 2) \emph{Access} \cite{azor}, and 3) \emph{LARC} \cite{larc}, compared to \emph{Oracle}. The hit ratio gap between \emph{Oracle} and state-of-the-art algorithms is up to 7x. This gap highly depends on the workload reuse distance and access pattern. Previous studies rely on such characteristics of workloads to identify hot data pages. In workloads with low localities, previous studies fail to predict which data pages will be re-referenced.

%%\begin{figure}[t]
%%\centering
%%\includegraphics[width=\linewidth]{motiv_oracle_hit.eps}
%%\caption{Oracle vs. Caching Algorithms: Hit Ratio} \label{fig:case_oracle_a}
%%\end{figure}
%%\begin{figure}[t]
%%\centering
%%\includegraphics[width=\linewidth]{motiv_oracle_rep.eps}
%%\caption{Oracle vs. Caching Algorithms: Replacements} \label{fig:case_oracle_b}
%%\end{figure}

In addition to the hit ratio, the number of cache replacements is also an important factor in SSD-based caching architectures. Cache replacements require a write operation on the SSD to copy the data page to the cache. Such write operations significantly decrease SSD lifetime and therefore, should be minimized. Figure \ref{fig:intro_motiv_b} depicts the number of cache replacements for the same set of workloads as the previous experiment. As shown in this figure, \emph{LARC} has the lowest number of cache replacements among state-of-the-art caching architectures. This is due to its two-level LRU, which prevents seldom accessed data pages from entering the cache. \emph{Oracle} has up to 13x less cache replacements compared to \emph{LARC}. Under \emph{cambridge} workload~\cite{cambridge}, where accesses have long reuse distance, previous methods such as \emph{LARC} fail to capture the relation between requests while most of the performance-critical blocks are evicted from the cache before they are accessed again. This results in significant hit ratio degradation compared to \emph{Oracle}. On the other hand, frequent swaps between blocks in the cache results in more cache replacements compared to the \emph{Oracle} caching architecture. The significant gap between the \emph{Oracle} and state-of-the-art cache architectures in terms of hit ratio and cache replacements reveal that caching architectures still can be significantly improved by employing more complex policies.

\subsection{Workload Characterization}\label{subsec:motiv_char}
In order to evaluate the accuracy of previously proposed characterization methods, three widely used characterization methods including a) \emph{Temporal Working Set Distribution} (TWSD) \cite{char_tarihi1}, b) Frequency \cite{azor, hystor}, and c) IOSize \cite{larc} have been implemented and examined. \emph{IOSize} characterizes requests based on their size. In \emph{Frequency}, requests are characterized based on the access frequency, type (read/write), and size. \emph{TWSD} method, which is the most detailed analysis on these three parameters (access frequency, type, and size), divides I/O requests into four types \emph{strided}, \emph{sequential}, \emph{random}, and \emph{overlapped} and takes the dependency of requests into account. A request is considered \emph{sequential}, if its size exceeds a threshold or it starts (ends) at the end (start) of one of previous requests. \emph{Strided} denotes requests with a small gap from previously accessed blocks. In addition, if a request overlaps with previous requests, it is flagged as \emph{overlapped}. Finally, if a request does not meet any of the mentioned conditions, it will be flagged as \emph{random}.

To evaluate the accuracy of the workload characterization methods, they are fed into an RNN model constructor as the main cost function. To be able to compare different workload characterization methods, we divide workload traces into two separate phases: \emph{learning} and \emph{evaluation}. During \emph{learning} phase, each method collects the required information from more than 16 trace files, each having at least 60,000 requests. Extracted information from the trace and the workload type are given to machine learning classifier model constructor to build a model for each workload characterization method. In the evaluation phase, each workload characterization method is tested 100 times, where each test consists of 100 requests. The accuracy is calculated by the number of correct workload identifications.

To simulate various storage servers with different workload complexities, three scenarios are designed using \emph{I/O traces} from \emph{SNIA} \cite{snia}. Table \ref{tab:scenarios} shows the three scenarios for the workloads.
\color{black}
In order to see the results of introducing one additional workload to a system, 
we decide to keep previous workloads exactly in later scenarios. Therefore, the reduced accuracy of characterization methods would not depend on the modified workload types. This experiment indicates the impact of additional workloads on the accuracy of characterization.
\color{black}
\emph{Single Purpose Server} data set emulates an email server by running an authentication application and an email server. \emph{Virtualization Server} runs three independent applications to emulate a server running several virtual machines. Finally, the last test data set (\emph{Storage System}) emulates a storage system by running several applications each having a separate filesystem.

\begin{table} 
\vspace{3mm}
\caption{Storage System Workload Scenarios} 
\begin{center} 
\vspace{-3mm}
\begin{tabular}[t]{| l | c |}
\hline
\textbf{Scenario} & \textbf{Workloads} \\ 
\hline
\emph{Single Purpose Server} & \begin{tabular}[x]{@{}c@{}}Radius\_Auth\\mail\_index\end{tabular}\\ 
\hline
\emph{Virtualization Server} & \begin{tabular}[x]{@{}c@{}}home\_ikki\\Radius\_Auth\\mail\_index\end{tabular}\\ 
\hline
\emph{Storage System} & \begin{tabular}[x]{@{}c@{}}enterprise\_tpc\_1\\home\_ikki\\Radius\_Auth\\mail\_index\end{tabular} \\ 
\hline
\end{tabular} 
\end{center} 
\label{tab:scenarios} 
\vspace{-6mm}
\end{table} 

\color{black} 
The goal of a characterization method is to find the type of running workload that the given set of I/Os belong to. To this end, the accuracy of characterization methods is calculated by the ratio of correct decisions of workload type over time, while workload types change periodically. For instance, in \emph{single purpose} scenario, characterization methods must only differentiate between two workload types~(Radius\_auth or mail\_index). Given one hundred  requests from a workload, the characterizer must decide which of the workload types are more likely to have generated the given hundred requests.
\color{black}

\begin{figure}[t]
\centering
\includegraphics[width=0.9\linewidth]{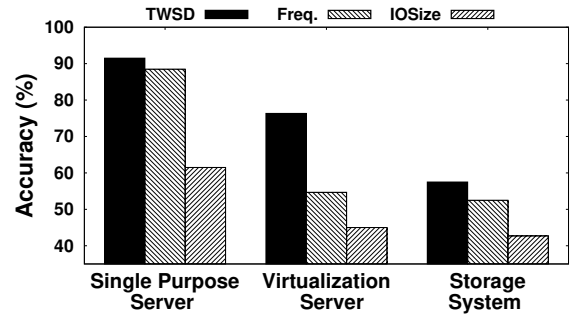}
\caption{Accuracy of Previous Workload Characterization Methods} \label{fig:motiv_char}
\vspace{-4mm}
\end{figure}

Figure \ref{fig:motiv_char} shows the accuracy of the workload characterization methods on test data sets. \emph{TWSD} and \emph{Frequency} have high accuracy when only one application is running. By increasing the number of running applications, the accuracy is reduced significantly. The highest accuracy in the third test data set is 60\%. This experiment reveals that the workload characterization methods fail to accurately predict workload when multiple applications are running. Therefore, more complex methods need be employed to reach high accuracy in workload characterization.

\vspace{-3.1mm}
\subsection{Exploiting Machine Learning Methods}\label{subsec:ml2}

\textcolor{black}{{Machine learning algorithms have shown to be successful in identifying complex data patterns in many domains by providing methods such as classification, regression, and clustering}}\cite{deep, ml2, ml4, ml5, label}\textcolor{black}{{. In addition, deep learning approaches such as RNNs are in general more accurate and more effective than the heuristics approaches that are currently employed in the caching management systems throughout the memory and I/O stack.
However, such machine learning approaches require few hundreds of microseconds for processing. The average latency of I/O requests is few milliseconds, which can tolerate such additional processing time.
However, main memory and CPU cache levels have less than one hundred nanoseconds latency, and therefore, cannot use machine learning approaches with such a \emph{relatively} high latency.}}

SSD caching, on the other hand, can benefit from machine learning and specially RNNs. This is due to three main reasons: 1) the significant gap between state-of-the-art architectures and Oracle, 2) RNNs as one of the most accurate discriminative classifiers for sequences with time dependency, and 3) the average response time of several milliseconds in I/O requests. Thus, RNNs can be employed to detect complex \emph{temporal} and \emph{spatial} localities of I/O workloads. We note that the average response time of I/O requests in this layer (several milliseconds) is sufficient for processing different RNN modules using only CPU. \textcolor{black}{{Moreover, other parameters such as request \emph{size} and \emph{type} (\emph{read}/\emph{write}) can be employed to further improve the accuracy of such methods by defining \emph{n}-dimensional neural vectors}}~\cite{andermatt2016multi}\textcolor{black}{{. Therefore, RNNs can be used for identifying patterns in long sequences with similar characteristics to I/O requests. Employing RNNs with LSTM}}~\cite{lstm} \textcolor{black}{{units can enable us to efficiently analyze long I/O workload traces. We can conclude here that I/O traces are suitable input for several machine learning methods such as RNN and the runtime computation overheads (as our experiments indicate) do \emph{not} have a significant effect on the overall performance.}}

\textcolor{black}{{Several studies have employed machine learning techniques in different levels of system hierarchy. C-Miner~}}\cite{cminer}\textcolor{black}{{ is a prefetching method for sequential prefetching in storage systems.
This method aims to provide faster I/O transactions between operating system and the backend storage system by considering the frequent subsequences in the blocks and \emph{pre-fetching} them to the memory.
This is orthogonal to RC-RNN that decides which data pages should be moved to the SSD \emph{after} they are accessed.
Additionally, the patterns that RNN can identify in the workload are not discoverable by simple mining methods such as \emph{frequent subsequences} that can only identify linear correlations, while I/O accesses have non-linear correlations.
Therefore, more complex machine learning approaches such as RNN should be employed to identify patterns in the I/O workloads.}}

\vspace{-2.5mm}

\section{Proposed Architecture}\label{sec:arc}

To mitigate the shortcomings of previous studies in terms of providing high hit ratio and SSD lifetime, we propose RC-RNN, the first reconfigurable SSD-based I/O caching architecture employing RNN models.
RC-RNN employs two RNN models: a) a single-layer RNN model for \emph{classification} of the running workload and matching its characteristics to one of the predefined workload categories, and b) a deep three-layer RNN model, called \emph{caching} model, to decide which data pages should be copied to/evicted from the cache.
For each workload category, a separate \emph{caching} RNN model is constructed, which is optimized toward the requirements of that specific workload type.
Since constructing RNN models is rather time-consuming, RC-RNN is divided into \emph{offline} and \emph{online} phases.
\textcolor{black}{{The offline phase is executed once for the entire lifetime of the system while the online phase monitors the I/O requests in the runtime. In the offline phase, RC-RNN constructs the required RNN models by analyzing a wide range of the workload traces from several enterprise applications. The fully built RNN models are later used in online phase to monitor workload and control the cache in real-time.}}
During online phase, the ongoing workload is periodically characterized to identify the workload type.
Based on the identified workload type, the corresponding RNN model is employed for cache management.
Fig. \ref{fig:arch} shows the detailed architecture and data flow of offline and online phases.
In the remainder of this section, offline and online phases are detailed.
\vspace{-2.5mm}
\subsection{Offline Phase} \label{subsec:offline}
\begin{figure}[t]%{0.55\linewidth}
\centering
\includegraphics[width=1.1\linewidth]{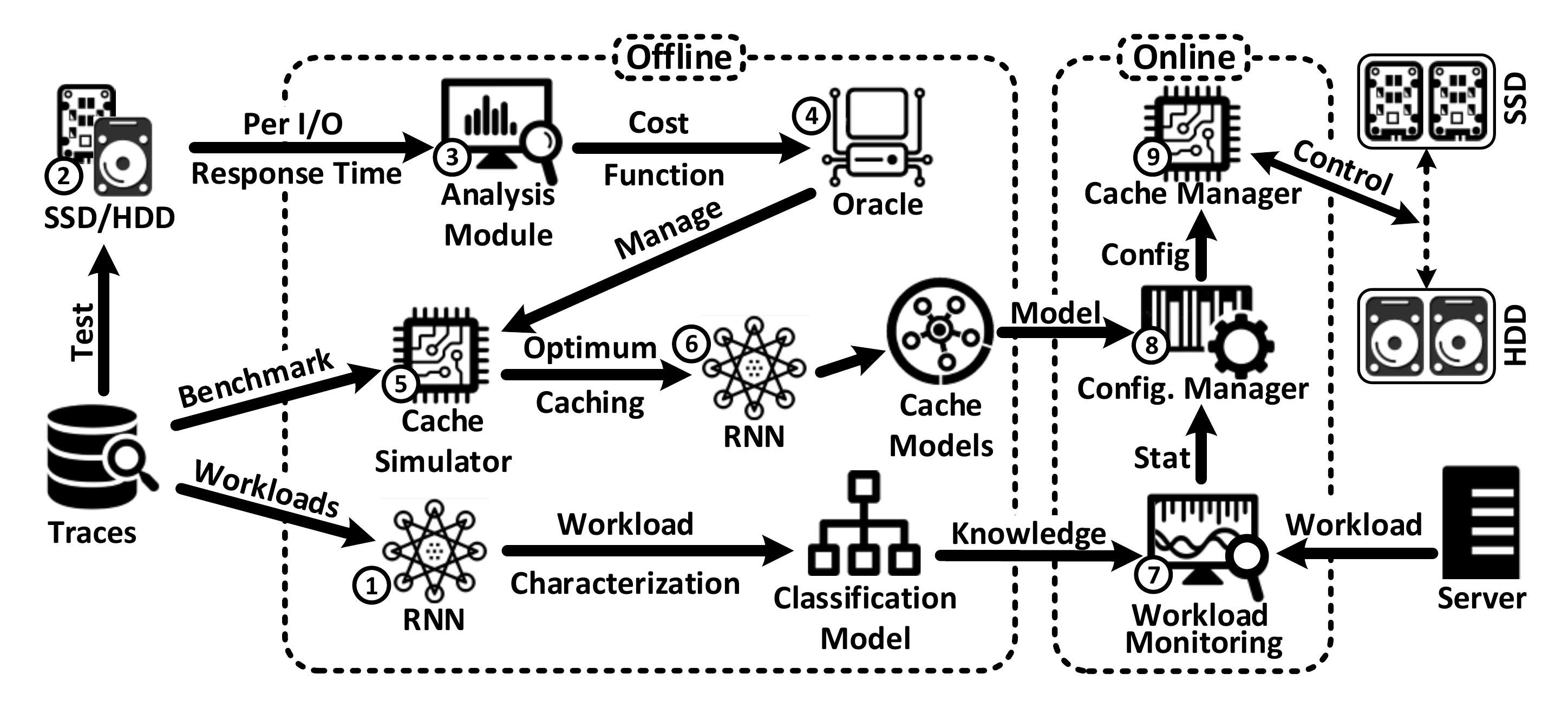}
\vspace{-3.5mm}
\caption{Overall Two-phase Architecture Flow}
\vspace{-3mm}
\label{fig:arch} 
\end{figure}

The offline phase is responsible for a) providing an RNN model for identifying workload type, and b) providing a cache management RNN model for each workload type. To identify workload types, a collection of I/O traces from SNIA \cite{snia} and \emph{filebench}~\cite{filebench} are examined and classified into four main workload categories: a) \emph{Mail Server}, b) \emph{Web Server}, c) \emph{Database}, and d) \emph{File Server}. The classification is based on the general trace categories in SNIA. As shown in Figure~\ref{fig:arch}, all workload traces and their corresponding type are fed into the RNN constructor \textcolor{black}{in a \emph{supervised} manner} \circled{1} to build a model for identifying workload type based on the I/O access pattern.

\color{black}
The characterization method in RC-RNN is based on the deep learning model. The characterization model learns the difference of given workloads in a supervised manner during the offline phase. Then, during online phase~(where evaluations are performed), the model decides which type of workload is more likely to have generated the requests in a given period of time. 
\color{black}

%\begin{figure*}[t]
%\centering
%\quad
%\caption{Proposed Architecture}\label{fig:arch_all}
%\end{figure*}

To construct RNN models for cache management, we need to label all requests by the decision of the Oracle.
The traditional Oracle algorithm \emph{only} considers the accesses to data pages in its cost function.
Due to the asymmetric performance of SSDs in read/write requests and also significant difference between sequential/random performance of HDDs, such algorithm might not result in the optimal cache decisions.
Therefore, we propose a benefit function, which can accurately estimate the benefit of caching data pages based on the several request characteristics.

Moreover, \circled{2} all of the traces are replied on real hardware (both SSD and HDD) to obtain the response time of the requests for each workload. The results are analyzed in \emph{Analysis Module} \circled{3} to calculate cost/benefit of caching each data page based on the access pattern. 
In addition to the response time of requests, read/write ratio and frequency of accesses to data pages are also considered in the benefit function.
The employed benefit formula for building RNN models of cache management is presented in Equation~\ref{eq:cost}.

As mentioned earlier in Section~\ref{sec:intro}, the performance of SSDs is highly depended on I/O request type~\cite{reca}. In order to use the most detailed SSD characterization analysis results, we employ \textcolor{black}{\emph{priority-based} caching mechanism with the benefit function~(equation~\ref{eq:cost}) to assign priorities to I/O requests. Therefore, requests with higher \emph{benefit} will have higher chance of being cached by the system. }
$T_{HDD}$ and $T_{SSD}$ denote HDD and SSD response times for a request, respectively. Requests having wider gap between SSD and HDD response times will benefit more from entering the cache and therefore, will be assigned higher priority.
Since requests with smaller sizes occupy less space in cache and larger requests have acceptable performance on HDDs, smaller requests assigned higher benefit by $\frac{1}{Req_{Size}}$ factor.

\begin{equation} \label{eq:cost}
Benefit = \frac{T_{HDD}}{T_{SSD}} \times (N_{acc} - 1) \times \frac{1}{Req_{size}} \times (1 + \frac{N_{reads}}{N_{acc}})
\end{equation}

\textcolor{black}{Improving read hit ratio reduces the number of data pages copied to the cache and therefore, reduces the number of writes in the SSD that increases SSD lifetime. Moreover,} SSDs provide higher performance in read requests contrary to HDDs, which have almost identical performance on read and write requests \cite{larc}. Therefore, benefit of caching read-intensive data pages should be higher than write-intensive data pages. To this end, the read ratio of accesses to the data page $(1 + \frac{N_{reads}}{N_{acc}})$ is also considered in the benefit function. In addition to improving performance, prioritizing caching of read-intensive data pages also improves SSD lifetime by reducing 
the number of writes in the SSD.

\emph{Oracle Cache} \circled{4} replays the trace file and for each access compares the benefit of caching a newly accessed data page with residing data pages in the cache. 
If the newly accessed data page has higher benefit, it replaces the data page with lowest benefit value in the cache.
\emph{Cache Simulator} \circled{5} is a simple cache, which manages the data structures for the cache based on the decisions of the \emph{Oracle Cache}.
For each access, \emph{Cache Simulator} adds two tags to the access in the trace file: a) \emph{cached} and b) \emph{duration}.
The \emph{cached} tag denotes that whether or not Oracle decided to place the data page in the cache.
The \emph{duration} tag denotes the number of accesses between entering a data page to the cache and replacing it with another data page.
The value of \emph{duration} depends on the cache size.
Since our RNN models need to be cache-size independent, the actual value of \emph{duration} is replaced with one of the three values reported in Table \ref{tab:thresh}.

The boundaries are selected based on the empirical studies and in such a way, three groups are almost uniformly populated. A three-level \emph{RNN} model \circled{6} learns the behavior of the Oracle by analyzing the tagged trace files created by \emph{Cache Simulator}.
Four RNN models are constructed, one for each workload category.
The output models are saved to be used later by the \emph{online} phase of RC-RNN.

\vspace{-2mm}
\subsection{Online Phase} \label{subsec:online}
\color{black}
{The RNN module tries to classify and distinguish four workload types by the five values provided for each I/O request.
As shown in previous works such as}~\cite{reca, char_tarihi2, char_tarihi1}, {the history of I/O accesses can be employed to predict the I/O behavior of applications.
Deep learning is able to identify the patterns in I/O workloads and its processing time is relatively small}~\cite{gheisari2017survey}.
{Hence, in this work we use deep learning to find the patterns in I/O workloads and decide the cache policy based on such information.}
\color{black}

\emph{Workload Monitoring} \circled{7} employs the RNN model for workload characterization and identifies the running workload type.
While the system is running, \emph{Workload Monitoring} is invoked once a minute.
It captures 1000 I/O requests and runs accesses through the RNN model for identifying workload type.
If a change in the workload type is detected, \emph{Workload Monitoring} replaces the current RNN model for cache management with RNN model of the new workload type.
The process of identifying workload type is done asynchronously to prevent any delay in responding to I/O accesses.

\emph{Configuration Manager} \circled{8} is responsible for reconfiguring the cache when a change in the workload is identified by \emph{Workload Monitoring}.
The reconfiguration process upon workload change consists of two stages, a) loading new RNN model, and b) re-evaluating benefit values of data pages in the cache based on the new configuration.
The reconfiguration process has very low overhead in terms of time and memory.
All cache models are generated during \emph{offline} phase and the \emph{Configuration Manager} only switches between models during workload change and responds to the newly arrived I/O requests using the
loaded model.
All models have the same computation time, which is less than one millisecond for each access, on average.
On multi-core CPUs, computation for I/O requests can be done simultaneously.
The average size of RNN models is less than 40 megabytes.
\color{black}
Note that the memory requirements of baseline SSD caching is about 0.2\% of the entire SSD capacity. For a sample SSD size of 1~TB, the baseline SSD caching requires at least 2~GB of memory. Therefore, the memory overhead of RC-RNN model is negligible. Therefore, all models can be loaded into the main memory during the cache startup. However, if the memory demand becomes significantly high, parts of lesser used RNN models can be swapped out by the virtual memory management of the OS.
\color{black}

\begin{table} 
\caption{Reclaiming Labels} 
\begin{center} 
\vspace{-3mm}
\begin{tabular}[t]{| l | c |}
\hline
\textbf{Label} & \textbf{Duration Areas} \\ 
\hline
\emph{soon} & $duration < \emph{Cache Size}$\\ 
\hline
\emph{mean} & $\emph{Cache Size}< duration < 5 \times \emph{Cache Size} $\\ 
\hline
\emph{late} & $5 \times \emph{Cache Size} < duration$\\ 
\hline
\end{tabular} 
\end{center} 
\label{tab:thresh} 
\vspace{-6mm}
\end{table} 

\begin{figure}[t]
\centering
\vspace{-5mm}
\includegraphics[width=\linewidth]{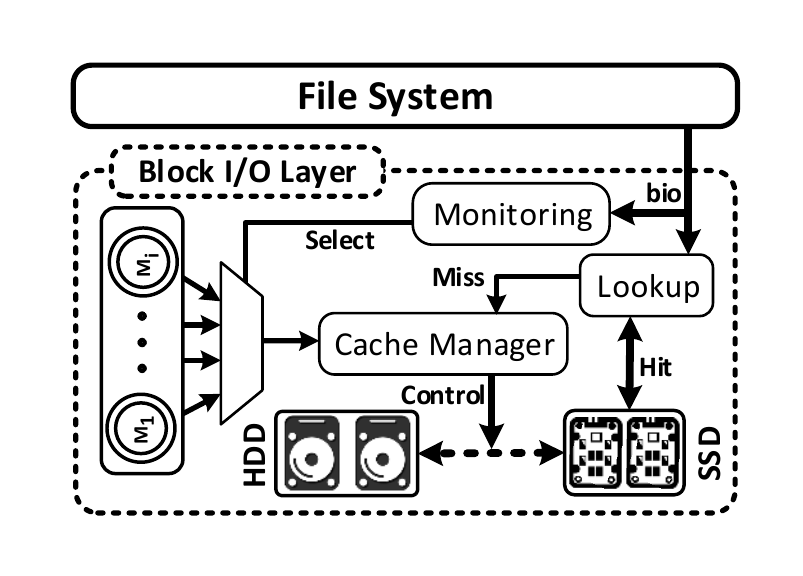}
\vspace{-10mm}
\caption{Online Phase Architecture} \label{fig:imp_arch}
\vspace{-5mm}
\end{figure}

Figure \ref{fig:imp_arch} shows the overall architecture of the implemented module for \emph{online} phase of RC-RNN.
\emph{Lookup} module maintains a list of data pages in cache and redirects hit accesses to the SSD.
Miss accesses are sent to the \emph{Cache manager} to decide whether or not it should be cached.
If the cache does not have any free space left, the cache manager decides which data pages should be evicted.
The \emph{Lookup} module handles all required I/O requests for moving data pages from/to the cache.
The \emph{Monitoring} module identifies current workload type and loads the corresponding RNN model into the \emph{Cache manager}.

\begin{figure}[t]
\centering
\includegraphics[width=\linewidth]{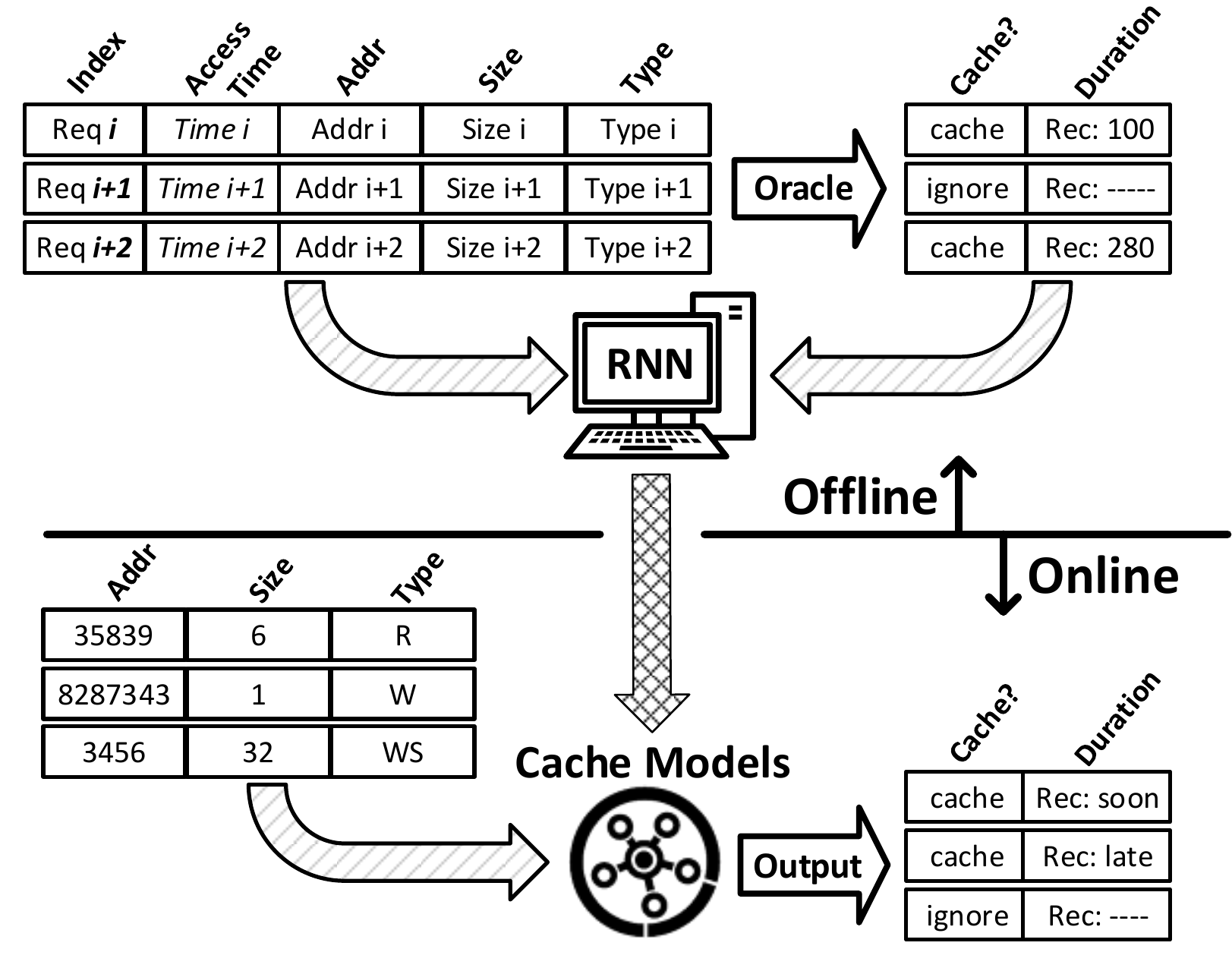}
\caption{Proposed Architecture and Dataflow for RC-RNN} \label{fig:oracle_rnn}
\vspace{-4mm}
\end{figure}

\emph{Cache manager} \circled{9} employs currently selected RNN model to decide a) whether or not to buffer the missed access, and b) which data page should be evicted when cache is full.
Fig. \ref{fig:oracle_rnn} shows the overall flow of decisions made by \emph{Cache manager} based on the outputs of Oracle.
The output of RNN model for each access directly states that the request should be buffered or not.
The eviction of data pages, however, requires additional process by the cache manager since RNN model only labels eviction time of requests as \emph{soon}, \emph{mean}, or \emph{late}.
To decide which pages should be evicted, RC-RNN maintains an LRU queue for each label and places requests in their corresponding queue.
On hit accesses, the requested data page will be moved to the head of its current queue.
Data pages are demoted to lower priority queues when $5 \times cachesize$ requests are processed after entrance of data pages to the cache.
Data pages in the \emph{mean} and \emph{late} queues are demoted to \emph{soon} and \emph{mean} queues, respectively.
To prevent demoting hot data pages, if a data page resides in top 20\% of the queue, it will not be demoted.
The victim for eviction will be the last data page in the \emph{soon} queue.
If the \emph{soon} queue is empty, the \emph{mean} and \emph{late} queues will be searched, respectively.
I/O caching architectures (unlike CPU caches or virtual memory) do not need to copy each missed access to the
cache and therefore, can bypass the cache and directly supplied by HDD.
This option enables RC-RNN to prevent seldom accessed data pages from entering the cache.

\color{black}
{Regarding general limitations of RNN such as the number of distinguishable classes, we note that RC-RNN does not try to predict the address of the upcoming requests.
During online workload characterization, RC-RNN predicts the workload category (e.g., Mail Server or File Server) based on the workload behavior.
The number of such categories is limited and at most four categories have been identified in previous studies}~\cite{reca,char_tarihi2,char_tarihi1}. {Moreover, based on the selected workload type, each I/O request is classified by a binary classification (to buffer or not to buffer) and if the model decides to buffer the data page, we use a further 3-class classification (\emph{soon}, \emph{mean}, and \emph{late}). Therefore, RC-RNN does not need to have a separate category for \emph{each} data page and the limitation of RNN in the number of categories does not affect RC-RNN.}
\color{black}
\vspace{-3mm}
\section{Experimental Results} \label{sec:exp_res}
In this section, first the detailed experimental setup for implementing and evaluating \emph{RC-RNN} is presented. Next, experimental results regarding tuning RNN models for both workload characterization and cache management are provided. We offer two scenarios for evaluation of \emph{RC-RNN}: 1) static workloads and 2) reconfiguration process. During each scenario, the accuracy of workload characterization model is evaluated. Afterwards, the proposed cache management model is compared to previous studies in terms of hit ratio and SSD lifetime. Finally, the impact of reconfiguration process is discussed while system is benchmarked under multiple workloads.

\vspace{-2mm}
\subsection{Experimental Setup}
\label{subsec:exp_stp}
\color{black}
\begin{table}
\centering
\caption{{Experimental Setup}}
\begin{tabular}{|c|c|} 
\hline 
\textbf{Device} & \textbf{Model} \\ 
\hline 
CPU & Core-i7 7500HQ \\ 
\hline 
Memory & 16 GB DDR3 \\ 
\hline 
HDD & 2 TB Western Digital Red Pro HDD \\ 
\hline 
SSD & 512 GB Samsung SSD 850 PRO \\ 
\hline 
GPU & NVIDIA TITAN X \textbf{(Only used during \emph{offline} phase)}\\
\hline 
\end{tabular} 
\vspace{-4mm}
\label{tab:exp_stp}
\end{table}
\color{black}

\emph{RC-RNN} is implemented as a kernel module based on \emph{EnhanceIO} \cite{enhanceio} in \emph{Linux Kernel 4.4.0}. A user-space application receives requests from kernel module, feeds them to the RNN model, and sends back the result to the kernel module. RNN models in the \emph{offline} phase are constructed by \emph{Keras} \cite{keras} library and \emph{RMSProp} \cite{rmsprop} optimizer with the default configurations. We consider two main hyperparameters to construct RNN models: a) number of hidden units, and b) I/O sequence window length. A system with 12 GB main memory with a \emph{NVIDIA Titan X} GPU running \emph{Ubuntu} 16.04 has been employed for constructing RNN models. \emph{Oracle} algorithm is implemented as a C++ module in an in-house cache simulator\footnote{The source code of simulator and RNN models will be publicly available upon acceptance of the paper.}. The complete log of the decisions made by \emph{Oracle} for every I/O request is stored to be used as an input for \emph{three-level deep} RNN model, which is shown in Figure \ref{fig:oracle_rnn}.

\color{black}
{Table}~\ref{tab:exp_stp} {presents the complete experimental setup for both offline and online phases of our evaluations. Note that the GPU is only utilized \emph{once} and \emph{only} in the offline phase.
In the experiments, the entire online evaluation is evaluated using \emph{only} CPU.
The online phase of RC-RNN in the experimental results uses the same hardware as LRU and LARC under all the workloads.
In all experiments, the size of the SSD is set to 20\% of the working set size of the workload. Table}~\ref{tab:traces} shows the number of requests, read/write ratio, and the average request size of traces.

% {Table}~\ref{tab:exp_stp}  {presents the complete experimental setup for both offline and online phases of our evaluations. Note that the GPU is only utilized \emph{once} and \emph{only} in the offline phase. The entire online evaluation is only evaluated using CPU and all of the caching methods utilize the same hardware under all of the workloads. The online phase of RC-RNN in the experimental results uses the \emph{same} hardware as LRU and LARC.
%The employed hardware for the experiments is reported in Table}~\ref{tab:traces} {.
%In all experiments, the size of the SSD is set to 20\% of the working set size of the workload.}
% In all experiments, the size of the SSD is set to 20\% of the working set size of the workload.}\textcolor{black}{ {Table}}~\ref{tab:traces} \textcolor{black}{ {shows the number of requests, read/write ratio, and the average request size of traces}}. 
{We note that the offline phase is not dependent to the running workload.
Therefore, as long as the overall hardware employed remains the same, we can reuse the results of the offline phase.
In case of a hardware change, the offline phase needs to be executed only \emph{once} before the start of the online phase.
In our experiments, the offline phase is executed for three hours and all the experimental results are captured during the online phase.
The results of online monitoring can be used as input traces for the offline phase.
This is an asynchronous operation and does not affect the performance of the online phase.
Although giving feedback from the online phase to the offline phase can result in more accuracy, it was not included in the results to have a fair comparison with previous architectures.}
\color{black}
\vspace{-3mm}
\subsection{RNN Configuration}
\label{subsec:exp_rnn}

%As mentioned in Section~\ref{sec:arc}, during the offline phase, \emph{Oracle}'s behavioral traces are fed to RNN models in order to 

\color{black}
The linear relation among I/Os has been extensively studied by the previous work, such as~\cite{reca, char_tarihi1, char_tarihi2}. In this paper, we aim to achieve higher accuracy in classifying the sequential data of I/Os by investigating the non-linear correlation in storage workloads. To this end, we used \emph{Recurrent Neural Networks} (RNN) that are designed to model sequential data. A usual RNN has a short-term memory, but in combination with  \emph{Long Short Term Memory} (LSTM), it will support a long-term memory. Moreover, different logistic activation functions, such as \emph{Softmax}, \emph{tanh} and \emph{ReLU} can be employed in implementation of RNN/LSTM. However, as we have multiple classes in our workload characterization, we choose \emph{Softmax} over other logistic alternatives. Finally, we utilize a three-layer network to reduce the overhead without sacrificing the desired performance.
\color{black}

\begin{table}[!t]
\centering
\caption{\textcolor{black}{{General Characteristics of Traces}}}
\begin{tabular}{|c|c|c|c|} 
\hline 
\textbf{Trace} & \textbf{\# of Reqs.} & \textbf{R/W Ratio} & \textbf{Avg. Req. Size} \\ 
\hline 
radius\_authentication &80k& 0.07 &12.44 KB\\
\hline radius\_backed\_SQL &60k& 0.21 &9.37 KB\\
\hline 
mail\_index &60k& 2.56 &42 KB\\ 
\hline 
home\_ikki &60k& 0.84&4 KB\\ 
\hline 
home\_madmax &60k& 0.25&4 KB\\ 
\hline 
home\_topgun &60k& 0.11&4 KB\\ 
\hline 
enterprise\_tpc\_1&90k& 2.05&8.24 KB\\ \hline 
MS\_enterprise\_ex&70k&0.16&21.6 KB\\\hline 
Cambridge1&65k&1.3&9.46 KB\\ \hline 
MS build server &60k&7.5& 4 KB\\ \hline 
MS live maps&70k&all read& 4 KB\\ \hline 
web proxy &60k&4.5&4 KB\\ \hline
web server &60k&12.7&4 KB\\
\hline 
\end{tabular} 
\label{tab:traces}
\vspace{-4mm}
\end{table}
\color{black}

The employed hyperparameters %(number of hidden units and I/O sequence window length) 
for constructing RNN models have significant impact on the accuracy of RNN models. Therefore, the optimal configuration for our RNN models should be selected before fully constructing the models. In order to determine the most efficient configuration for implemented RNN, we test the models with multiple number of hidden units under different workload characterization scenarios described in Table~\ref{tab:scenarios}. After tuning the configurations in the workload characterization of the RNN model, evaluations are done in comparison against \emph{Oracle} algorithm.

\subsubsection{Workload Characterization Model}\label{subsec:exp_wrk_char_rnn}
The input of the one-level RNN model for characterization is sequences of 100 consecutive requests containing four data fields: (1) \emph{arrival time}, (2) \emph{address}, (3) \emph{size} and (4) \emph{request type}, which are detailed in Figure \ref{fig:oracle_rnn}. The output is a \emph{softmax} layer containing the estimated probabilities of each data field for workload types.

\begin{figure}[t]%{\linewidth}
\centering
\includegraphics[width=\linewidth]{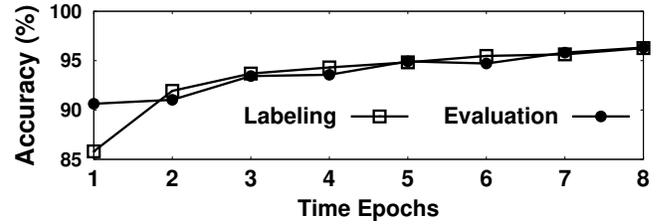}
\caption{Accuracy in Learning Phase of Proposed Characterization Method} \label{fig:exp_char_a}
\vspace{-4mm}
\end{figure}

\emph{Learning} phase consists of two repetitive actions: 1) \emph{labeling} and 2) \emph{evaluation}. RNN (re)labels the input data sequences and evaluates its labeling by predicting the test data. In the next iteration, RNN corrects itself to reach higher accuracy in classifying data. {In order to evaluate the accuracy of the workload characterization model, a part of the workload sequences are used for constructing RNN models in the \emph{learning} phase and entirely different sequences are employed for testing the accuracy of the proposed RNN model.} Figure \ref{fig:exp_char_a} shows the accuracy of \emph{labeling} and \emph{evaluation} of the proposed RNN-based characterization method during \emph{learning} phase of \emph{Storage System} workload (listed in Table~\ref{tab:scenarios}). In this scenario, the accuracy converges into 95\% after 8 iterations where it saturates. Once the model reaches its saturation, learning phase is completed. {We note that all of samples from employed I/O traces are cross-validated over a large sequence of available I/O workloads to make sure practicality of trained RNN models in general.}

\begin{figure}[t]%{\linewidth}
\centering
\includegraphics[width=1.1\linewidth]{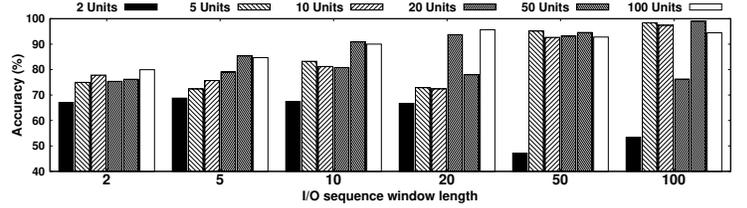}
\caption{\textcolor{black}{{Accuracy of Implemented RNN with Different Configurations for Proposed Characterization Method}}} \label{fig:exp_rnn}
\vspace{-2mm}
\end{figure}

Figure \ref{fig:exp_rnn} shows the accuracy of various RNN configurations under the first scenario (\emph{Single Purpose Server}) of Table~\ref{tab:scenarios}. \textcolor{black}{{Experiments reveal that in most cases, increasing I/O sequence window length results in higher accuracy as RNN has longer sequences and can build deeper and more detailed insight to learn the data access pattern of workloads}}. Another observation is that increasing the number of hidden units in RNN structure will not always result in higher accuracy. For instance, 100 hidden units configuration has lower accuracy than 50 hidden units configuration using a sequence window size of 100. This is due to \emph{saturation} of model with 50 hidden units as the accuracy reaches above 99\% and the model learns the input pattern completely. In this situation, higher structure overheads such as longer window length and higher number of hidden units will not result in higher accuracy. Based on such observations, we configure RNN models with 50 hidden units and window size of 100 I/O requests.
\color{black}
Note that the single layer RNN model already achieves more than 99\% accuracy. Therefore, adding extra layers to the model would only result in a more complex model without accuracy gain.
\color{black}

%\vspace{-1mm}
\begin{table}[t] 
\caption{Optimizing Learning of the Oracle Using RNN} 
\begin{center} 
\vspace{-7mm}
\begin{tabular}[t]{ | l | c | c | c | c | c | c | c | }
\hline
\multirow{2}{*}{\begin{tabular}[x]{@{}c@{}}Workload\\Name\end{tabular}} & \multicolumn{2}{c|}{RC-RNN} & \multicolumn{2}{c|}{Baseline-RNN} & \multirow{2}{*} {\begin{tabular}[x]{@{}c@{}}RCRNN\\imprv.\end{tabular}}\\
\cline{2-5}
& cached & durat. & cached & durat. & \\ \hline
MS ent. ex. & 91.23\% & 93.65\% & 62.69\% & 47.74\% & 185\%\\
\hline
cambridge1 & 97.93\% & 95.17\% & 72.63\% & 60.34\% & 112\%\\
\hline
home ikki & 92.34\% & 100.00\% & 73.42\% & 54.76\% & 129\%\\
\hline 
home topgun & 97.59\% & 94.90\% & 78.23\% & 65.52\% & 80\%\\
\hline 
home mdmx. & 98.35\% & 92.43\% & 71.61\% & 58.13\% & 118\%\\
\hline 
Radius Auth. & 89.31\% & 91.42\% & 58.40\% & 42.92\% & 225\%\\
\hline 
Radius b. sql &95.45\% & 72.54\% & 66.62\% & 43.49\% & 138\%\\
\hline 
ent. tpcE & 88.31\% & 100.00\% & 83.31\% & 74.18\% & 43\%\\
\hline 
mail index & 94.32\% & 99.53\% & 76.98\% & 90.71\% & 34\%\\
\hline 
MS bld. serv. & 97.59\% & 99.90\% & 77.60\% & 72.83\% & 73\%\\
\hline 
MS live map & 93.74\% & 100.00\% & 82.75\% & 98.63\% & 15\%\\
\hline 
msn data & 97.17\% & 91.55\% & 78.14\% & 61.54\% & 85\%\\
\hline 
msn m.data & 94.99\% & 97.07\% & 75.40\% & 68.99\% & 77\%\\
\hline 
varmail & 96.74\% & 98.44\% & 63.47\% & 55.17\% & 172\%\\
\hline 
web proxy & 93.51\% & 99.67\% & 71.20\% & 52.87\% & 147\%\\
\hline 
web server & 95.23\% & 98.50\% & 78.36\% & 68.55\% & 75\%\\
\hline
\end{tabular} 
\end{center} 
\label{tab:rnn} 
\vspace{-4mm}
\end{table} 

\subsubsection{Caching Management Model}

In order to construct the RNN model for cache management, a three-level RNN is employed. Each level in the RNN model consists of 256 LSTM units. Workloads are divided into sequences of 100 consecutive requests. Each request is a five dimensional vector of \emph{address}, \emph{size}, \emph{request type}, \emph{cached/ignored}, and \emph{duration} label, which have been presented in Section \ref{sec:arc}. The output of the RNN for each workload type is a model, which is able to mimic \emph{Oracle} decisions under the running workload. By setting batch size to 32 and optimizer to \emph{RMSProp} \cite{rmsprop}, we train the model and report the accuracy of decision for the workload, independently.

\color{black}
{The contribution of this work is not limited to choosing RNN model.
The effect of several novel optimizations employed in this work is shown in Table}~\ref{tab:rnn}.
{\emph{Baseline-RNN} denotes the accuracy of an RNN model without the proposed cost/benefit function.
RC-RNN significantly improves the accuracy of the baseline RNN by employing a novel cost/benefit function, which considers the performance of both HDD and SSD.
Additionally, RNN can \emph{only} predict the behavior for limited number of categories.
Therefore, predicting the exact data page for eviction is not possible by using RNN.
However, RC-RNN classifies data pages into three categories (\emph{soon}, \emph{mean}, and \emph{late}) in order to use RNN, while maintaining a high accuracy.}
\color{black}

Table \ref{tab:rnn} shows the accuracy of the proposed model in deciding \emph{cached} and \emph{duration} tags for requests. In addition to the proposed model, the accuracy of a baseline RNN model without the proposed cost function has also been evaluated and presented in Table~\ref{tab:rnn}. The \emph{baseline} model selects the most frequent tag value and assigns it to all requests for each iteration. For instance, if \emph{Oracle} tags more than 50\% of requests as \emph{cached}, the static model tags all requests in the evaluation phase as \emph{cached} and vice versa. Under workloads such as \emph{enterprise tpcE} and \emph{MS live maps}, the \emph{Oracle} algorithm tags more than 80\% of the requests as \emph{ignored}. The \emph{basic} classifier tags all requests as \emph{ignored} and therefore, has above 80\% accuracy.

\emph{Baseline} classifier shows the maximum accuracy achievable by static methods. The improvement of \emph{RC-RNN} compared to \emph{basic} classifier is calculated based on Equation~\ref{eq:rnn_static} and is shown in the last column of Table~\ref{tab:rnn}. Overall, the cost function provided by RC-RNN improves the accuracy by more than 106\% on average, which shows the significance of defining the cost/benefit function.
%\vspace{-0.1mm}
\begin{equation}
\label{eq:rnn_static}
RCRNN\_Imprv. = \frac{RCRNN_{cached} \times RCRNN_{dur.}}{baseln_{cached.} \times baseln_{dur.}}
\end{equation}
%\vspace{2mm}
%\subsubsection{Static Scenario}
\color{black}
\vspace{-3mm}
\color{black}
\subsubsection{RNN models overhead}
{To show the overhead of processing RNN model in runtime, we measure the latency of the RNN decision-making process (during the online phase and \emph{without} using GPU).
Using 400,000 (1000$\times$[100$\times$4]) samples with a batch size of 32, RC-RNN adds only 0.3ms overhead to each processed I/O request.
This adds only 15\% overhead to the response time of the requests.
Processing of parallel I/O requests can be done simultaneously, which further reduces the actual response time overhead of RC-RNN.
In the experiments, the average response time overhead is less than 10\%.} %{Although absolute latency seems to be high for NVMe SSDs, the increase in the cache hit ratio by employing RC-RNN compensates the processing overhead.}

\color{black}
Although the overhead of RC-RNN might be relatively high for low-latency SSDs, it still reduces the average response time of IOs when low-latency SSDs are employed. This is because RC-RNN significantly increases the hit ratio of the cache. Therefore, it issues less requests to HDDs, which are typically much slower than SSDs. Since the response time of HDDs is in the range of several milliseconds, each access to HDDs that is serviced by the cache will overcompensate the CPU overhead imposed by running RC-RNN. This is shown in the experimental results of Section~5.3.
The benefit of reducing HDD accesses not only compensates the overhead of RC-RNN, but also reduces the average latency. Additionally, as is shown in Section~5.3, RC-RNN requires much less cache replacements on average, which again reduces the number of accesses to HDDs.

Finally, we note that the overhead of RC-RNN is not dependent on the storage capacity or IOPS. Therefore, it can be employed on top of storage systems with tens of terabytes of storage. In addition, the memory overhead of RC-RNN is negligible compared to the overhead of conventional SSD caching. For instance, RC-RNN adds only 8\% additional memory overhead on top of the memory requirements for basic SSD caching with the size of 1TB.
\color{black}

% {We note that, we have reached the 0.3ms delay by using only one general purpose CPU. Therefore, when the 0.3ms delay causes the significant overhead to the system, we can employ more and faster CPUs in order to decrease the calculation delay. Moreover, for more high-end storage systems that utilize faster SSDs, one can benefit from employing a dedicated GPU or DNN ASIC co-processors to accelerate decision process of RC-RNN.}

\subsection{Static workloads Scenario}
\label{subsec:exp_wrk_char}

In this section, we evaluate both proposed characterization and cache management methods against previous studies in a scenario where workloads are static. The results present the performance of each method in a long-term single-purpose environment. At first, we evaluate proposed workload characterization method against state-of-the-art characterization methods. Finally, the proposed caching architecture is compared to previous studies in terms of hit ratio and SSD lifetime. Experiments in this section are done by employing a static scenario where there is no change in the running workload. To evaluate caching architectures, 16 traces from \emph{SNIA}~\cite{snia} and \emph{Filebench}~\cite{filebench} have been selected. The proposed model is compared to \textcolor{black}{four} widely used caching architectures: (1) \emph{LRU}, (2) \emph{Access} \cite{azor, hystor} (3) \emph{LARC}~\cite{larc}\color{black}, and~(4)~\emph{RECA}~\cite{reca}\color{black}. In addition, \emph{Oracle} has been implemented to show the maximum possible hit ratio and SSD endurance efficiency. 

\begin{figure}[t]%{\linewidth}
\centering
\includegraphics[width=0.75\linewidth]{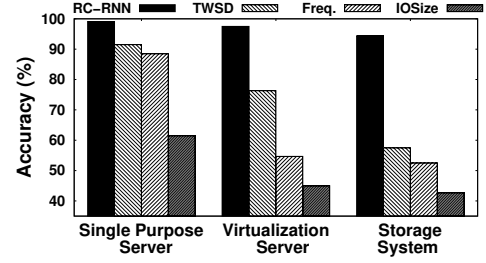}
\caption{Accuracy of Proposed Characterization Method Compared to Previous Studies} \label{fig:exp_char_b}
\vspace{-4mm}
\end{figure}

We evaluate the proposed characterization method in three scenarios presented in Table \ref{tab:scenarios}. Figure \ref{fig:exp_char_b} shows the accuracy of the proposed characterization method compared to the previous studies. The proposed method has an accuracy of 99\% in \emph{Single Purpose Server} scenario. The results show that in \emph{Virtualization Server} and \emph{Storage System} scenarios, previous workload characterization methods fail to accurately predict the workload, while the proposed model is able to accurately identify workload type with more than 94\% accuracy. The high accuracy of the proposed model (up to 55\% higher than previous studies) enables \emph{RC-RNN} to identify workload type in almost all scenarios.

\begin{figure}[t]%{\linewidth}
\centering
\includegraphics[width=\linewidth]{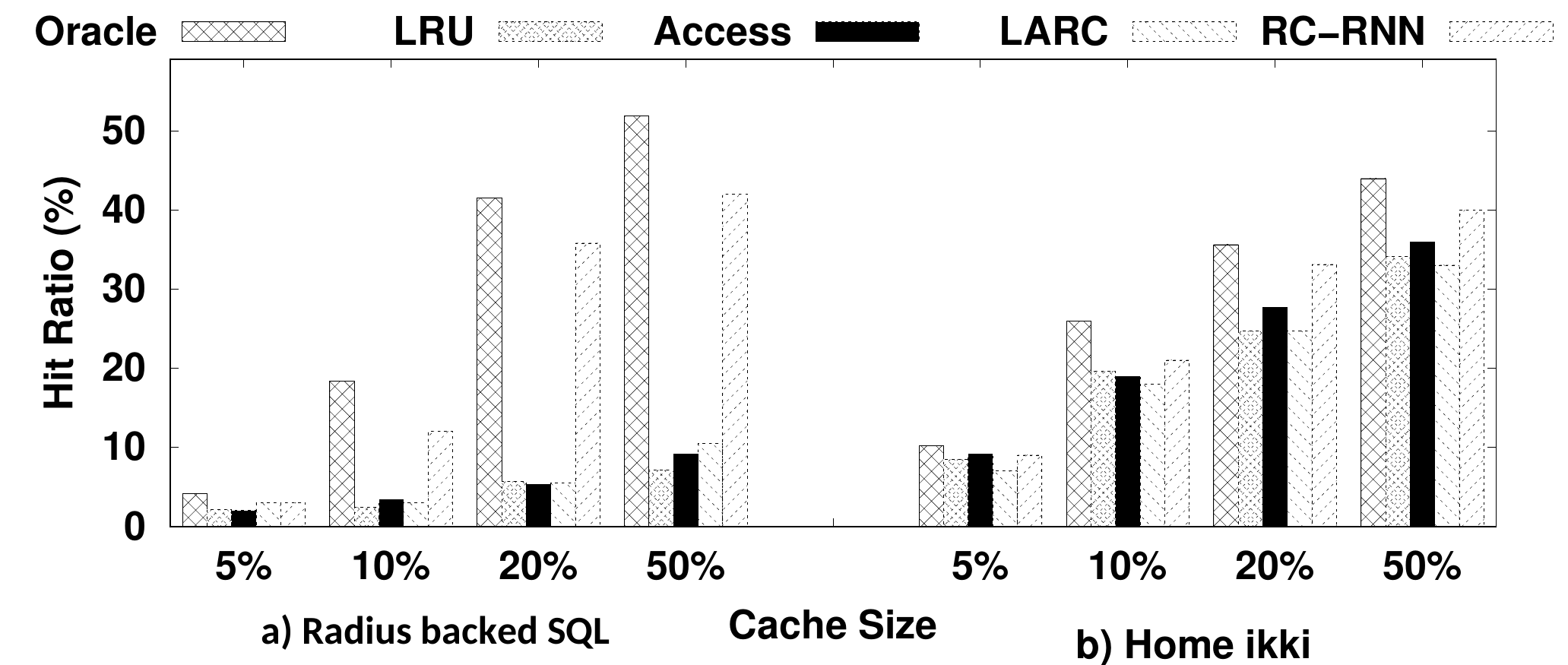}
\caption{ {Impact of Cache Size on the Overall Hit Ratio}}
\vspace{-4mm}
\label{fig:exp_cch_size}
\end{figure}

\color{black}
{Larger caches can hold more data pages and therefore, have a higher hit probability.
The cache hit ratio, however, does not increase linearly with the cache capacity.
To show the effect of cache size on the hit ratio, we experiment several cache sizes for two workloads, shown in Fig.}~\ref{fig:exp_cch_size}.
 {We selected Radius and Home workloads which represent other workloads in this case. All omitted workloads behave similarly to one of the two presented ones.}
Employing SSD caches with a size of more than 20\% of the size of working set size is not economically justified, since the cost of the SSD outweighs the performance improvement.
Thus, we use 20\% of the working set size as the cache size in our experiments.
In our analysis, most of the cache misses are because of insufficient capacity. Cache misses due to the limited associativity are negligible in most workloads. Additionally, we measure the cache misses after the SSD is filled, and therefore, cold cache misses are not considered in our evaluations.

\begin{figure*}[thp]
\centering
%\begin{subfigure}[t]{\linewidth}
\includegraphics[width=\linewidth]{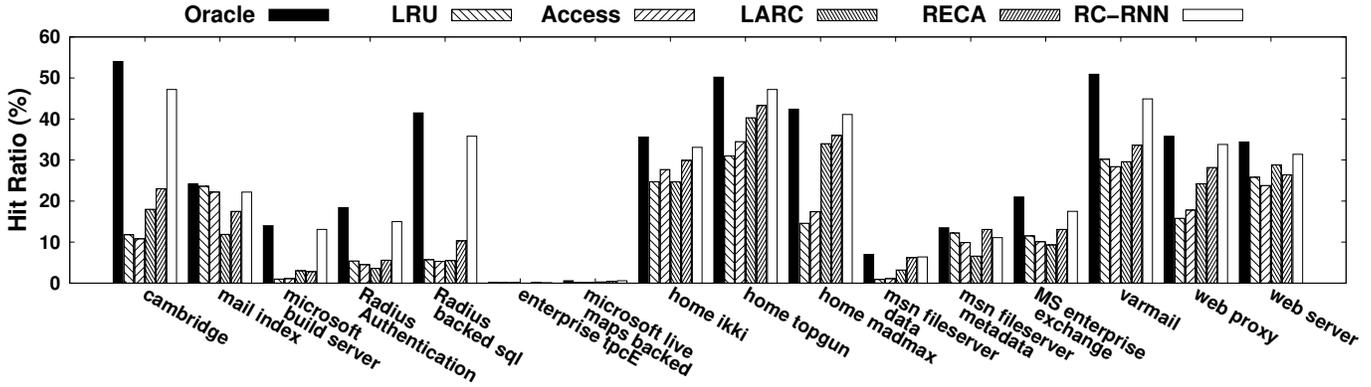}
\vspace{-6mm}
\caption{\textcolor{black}{Cache Performance (Hit Ratio)}} \label{fig:exp_cch_static_a}
\end{figure*}

\begin{figure*}[thp]
\centering
%\begin{subfigure}[t]{\linewidth}
\includegraphics[width=\linewidth]{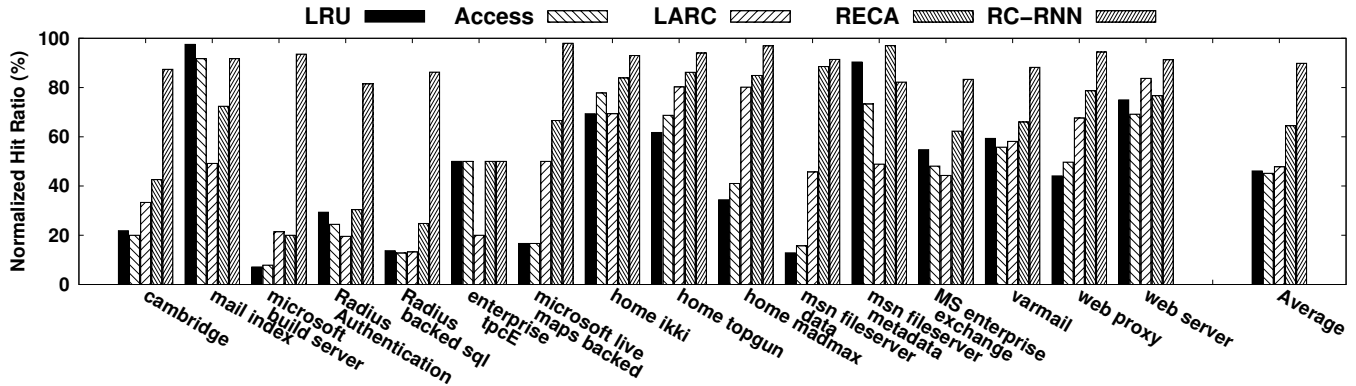}
\vspace{-6mm}
\caption{\textcolor{black}{Normalized Hit Ratio of Different Cache Architectures Over Oracle}} \label{fig:exp_cch_static_c}
%\end{subfigure}
%\caption{Static Evaluation Scenario}\label{fig:exp_cch_static}
\end{figure*}

\begin{figure*}[t]
\centering
%\begin{subfigure}[t]{\linewidth}
\includegraphics[width=\linewidth]{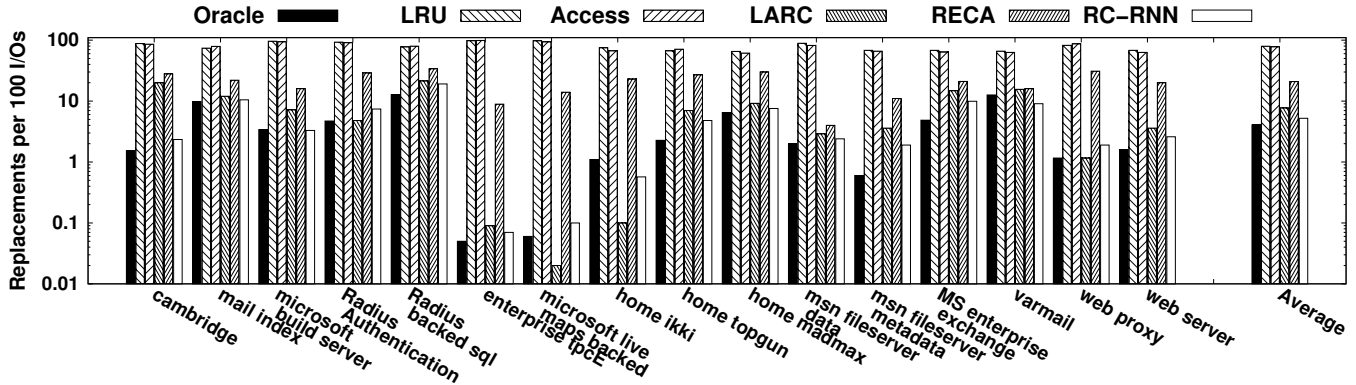}
\vspace{-6mm}
\caption{\textcolor{black}{Number of Cache Replacements}} \label{fig:exp_cch_static_b}
%\end{subfigure}
%\caption{Static Evaluation Scenario}\label{fig:exp_cch_static}
\vspace{-4mm}
\end{figure*} 

\begin{figure*}[t]
\centering
%\begin{subfigure}[t]{\linewidth}
\includegraphics[width=\linewidth]{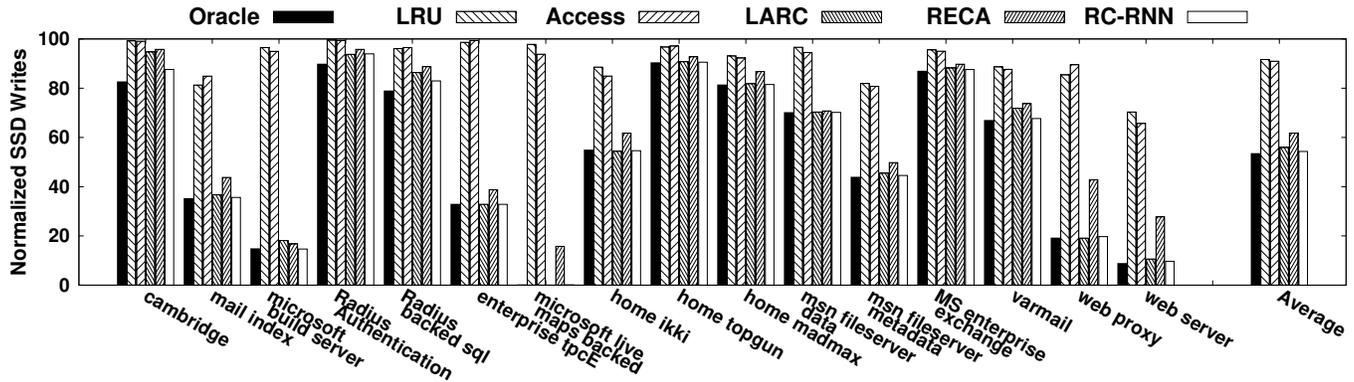}
\vspace{-6mm}
\caption{ {Average Number of SSD Writes per 100 I/Os}} \label{fig:exp_cch_static_wr}
%\end{subfigure}
%\caption{Static Evaluation Scenario}\label{fig:exp_cch_static}
\end{figure*} 

\begin{figure*}[t]
\centering
%\begin{subfigure}[t]{\linewidth}
\includegraphics[width=\linewidth]{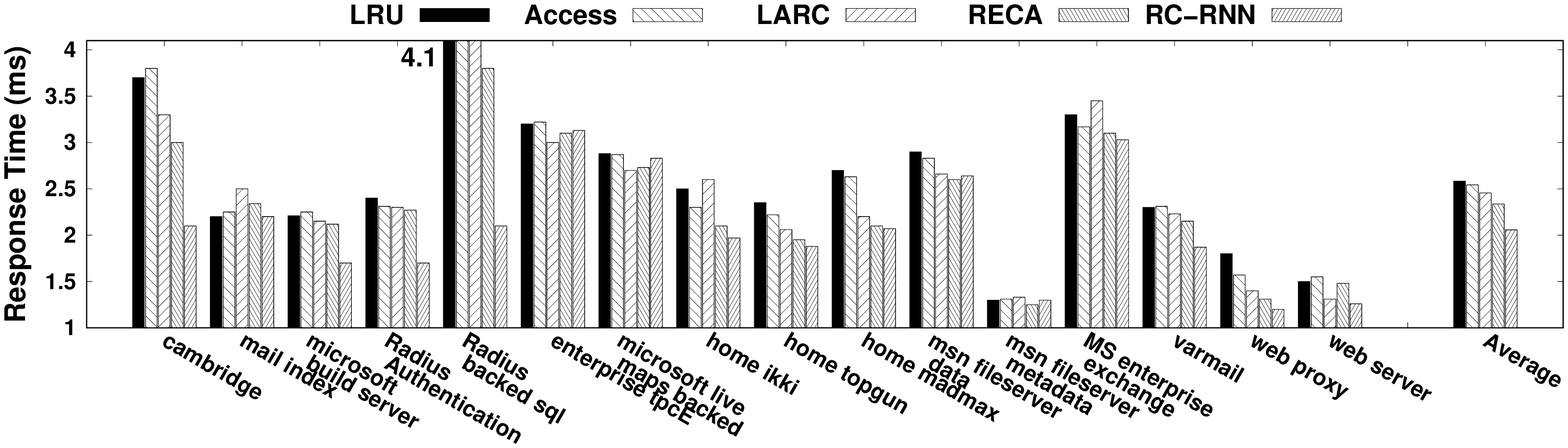}
\vspace{-6mm}
\caption{\textcolor{black}{Overall System Performance (Average I/O Response Time)}} \label{fig:exp_cch_static_res}
%\end{subfigure}
%\caption{Static Evaluation Scenario}\label{fig:exp_cch_static}
\vspace{-4mm}
\end{figure*}

Figure \ref{fig:exp_cch_static_a} shows the hit ratio of several caching architectures under various workloads. \emph{RC-RNN} performs 95\% similar to \emph{Oracle} on average and therefore, outperforms previous architectures with up to 7x higher hit ratio in workloads such as \emph{Radius\_backed\_sql} and \emph{Cambridge}. The hit ratio gap between \emph{Oracle} and state-of-the-art caching architectures highly depends on the workload type. Previous studies manage cache based on \emph{recency} and \emph{frequency} of the I/O requests. Therefore, in workloads such as \emph{main index} or \emph{MSN fileserver metadata}, which have high linear temporal and spatial localities, the hit ratio of previous studies is close to \emph{Oracle}. For most of the workloads, however, the hit ratio gap between \emph{Oracle} and previous studies is considerably high. \emph{RC-RNN} on the other hand, is able to achieve high hit ratio in most workloads due to the proposed RNN-based cache management model. The normalized hit ratio values of Figure \ref{fig:exp_cch_static_a} compared to \emph{Oracle} are shown in Figure \ref{fig:exp_cch_static_c}. Previous studies have less than 63\% of the maximum possible hit ratio. \emph{RC-RNN} has up to 90\% of the maximum possible hit ratio, which is up to 30\% higher.% than previous studies.

In addition to hit ratio, the number of cache replacements is also analyzed. Reducing the number of cache replacements improves SSD lifetime and therefore reduces the cost of device replacements.
 {Figure~}\ref{fig:exp_cch_static_b}  {presents normalized cache replacements per 100 I/Os for each caching architecture.}
\emph{RC-RNN}, while having superior hit ratio compared to previous studies, lowers cache replacements in several workloads such as \emph{cambridge} and \emph{MS Build Server}. The number of cache replacements is reduced by up to 8x compared to \emph{LARC}.
In \emph{Radius Authentication} workload, \emph{RC-RNN} and \emph{LARC} have almost the same number of cache replacements.
The hit ratio of \emph{RC-RNN}, however, is 7x higher than \emph{LARC}. This is due to selecting more suitable data pages for caching in \emph{RC-RNN}. In \emph{cambridge} workload, \emph{RC-RNN} has 8x less cache replacements while achieving more than 2.7x higher hit ratio compared to \emph{LARC}. Our analysis shows that in this workload, working set size is larger than cache size. Therefore, \emph{LARC} cannot benefit from data pages in cache while \emph{RC-RNN} can detect the performance-critical data pages and keeps such data pages in cache till they are re-referenced. %\emph{RC-RNN} does not fetch data pages from HDD to SSD if they are not requested by the applications.
%Additionally, it may decide to skip buffering a data page, which also results in not fetching the page to the SSD.
%Therefore, the number of fetches in RC-RNN is lower than the state-of-the-art I/O caching architectures.

 {Figure~}\ref{fig:exp_cch_static_wr}  {shows the number of SSD writes per 100 I/Os. In almost all workloads, RC-RNN has a less or equal number of SSD writes, compared to LARC and RECA. This shows that RC-RNN not only improves the performance but also extends the SSD lifetime. Our analysis shows that most of the SSD lifetime improvement originates from the reduction in cache replacements.}

Figure~\ref{fig:exp_cch_static_res} shows the average response time of caching mechanisms under various workloads. RC-RNN reduces the response time of standard benchmarks by 15\% on average in comparison with the best of the previous work~(RECA~\cite{reca}). The RC-RNN's computational overhead is only noticeable when the caching mechanism fails to deliver high hit ratios in workloads such as \emph{``enterprise TPC-E''} and \emph{``MS live maps backup''}, where \emph{none} of the caching mechanisms provide any improvement. Other than that, in most of the scenarios, RC-RNN reduces the average response time of the workloads by up to 1.6 ms~(43\%). The computational overhead of RC-RNN~(0.3 ms) has been considered in our performance measurements. These results suggest that employing RNN for SSD-based caches in storage systems can be beneficial in cases where the state-of-the-art statistical caching mechanisms fail to deliver acceptable cache performance in comparison with \emph{Oracle}. Such scenarios are especially noticeable under workloads such as \emph{cambridge}, \emph{MS build server}, \emph{Radius backed\_sql}, and \emph{Radius authentication}.

\begin{figure}[t]%{\linewidth}
\centering
\includegraphics[width=0.9\linewidth]{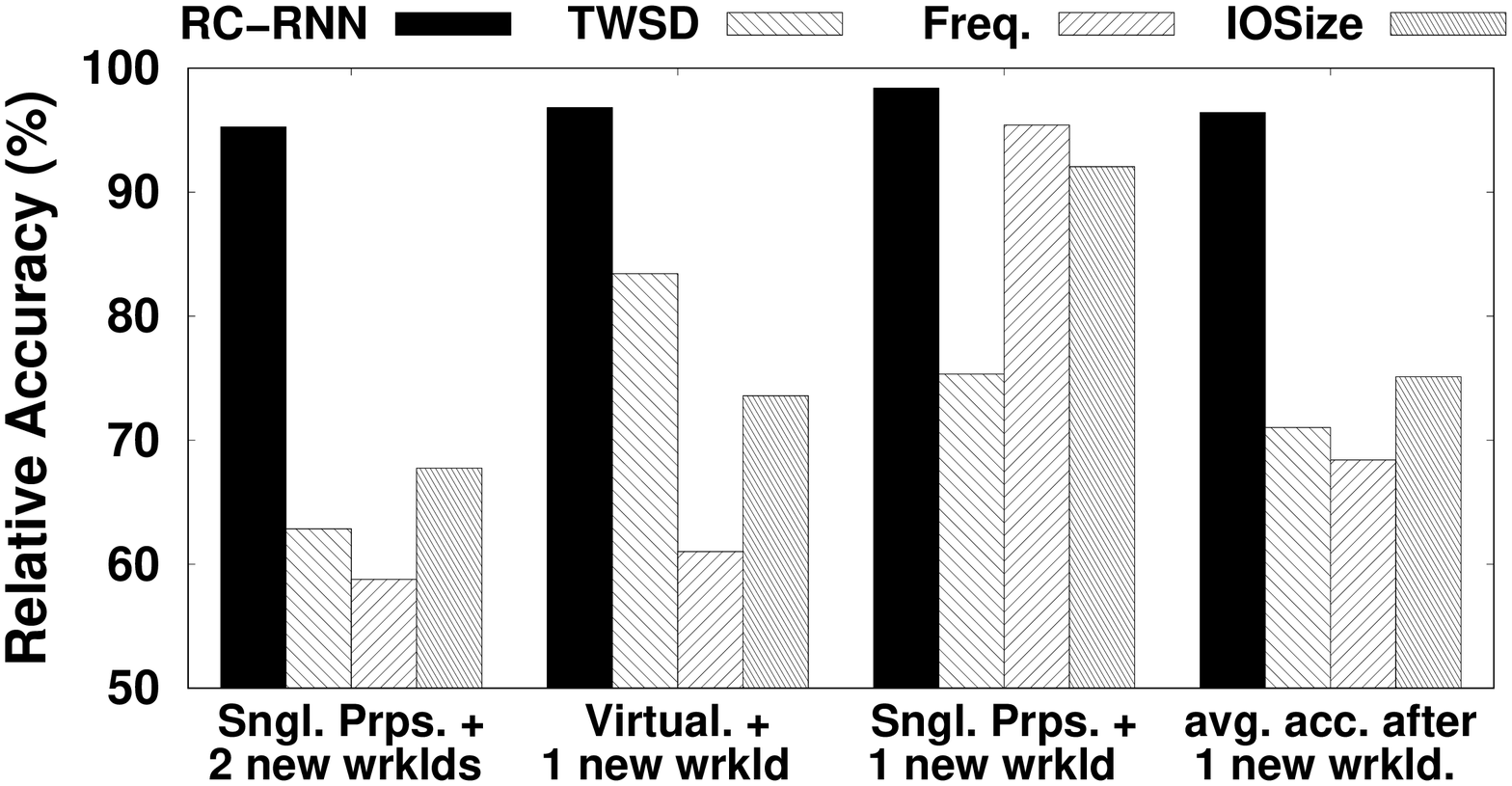}\vspace{-3mm}
\caption{Relative Accuracy of Characterization Methods upon Addition of New Workloads} \label{fig:exp_char_c}
\vspace{-6mm}
\end{figure}

\vspace{-2mm}
\subsection{Reconfiguration Process}
\label{subsec:exp_cch_mngmnt}
In this section, the proposed characterization method and cache management architecture are evaluated in case of multiple servers running several applications to simulate the behavior of enterprise storage systems. First, we compare the accuracy of the proposed RNN-based characterization method against previous work. Later, the impact of reconfigurability of the proposed architecture using multiple RNN models for cache management is analyzed. The most important parameter of workload characterization methods is the accuracy of identifying workload type when a new application starts. Workload characterization methods lose accuracy upon adding an application to the running applications. Methods having less accuracy loss are considered to be more efficient. Figure \ref{fig:exp_char_c} shows the accuracy of various workload characterization methods upon adding a new application. While previous studies have up to 40\% accuracy loss, \emph{RC-RNN} maintains its accuracy with less than 5\% accuracy loss.
As shown in Figure~\ref{fig:exp_char_c}, for workload characterization RC-RNN can achieve magnificently higher accuracy (up to 30\%) than TWSD~(the proposed method in~\cite{reca}).

\begin{figure}[t]
\centering
\includegraphics[width=\linewidth]{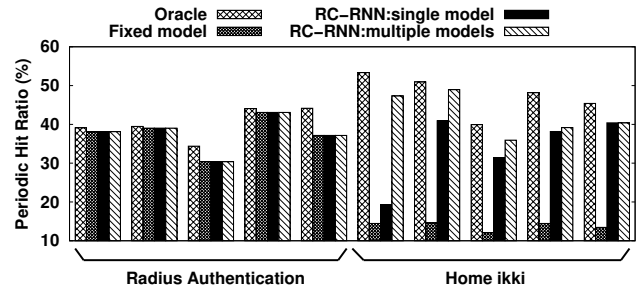}\vspace{-2mm}
\caption{\textcolor{black}{Periodic hit ratio in case of workload change}}\label{fig:exp_cch_dyn}\vspace{-4mm}
\end{figure}

To evaluate the impact of reconfiguration process on the overall performance of \emph{RC-RNN}, hit ratio is observed in case of a workload change. The workload consists of 50,000 requests from \emph{radius\_authentication} followed by 50,000 requests from \emph{home\_ikki} workload. Figure \ref{fig:exp_cch_dyn} shows the performance of \emph{RC-RNN} with and without considering the reconfiguration.
\emph{Fixed\_model} has \emph{only} one model for all workload types and \emph{RC-RNN} employs a model for each workload type.
\color{black}
In addition, we show the reconfiguration overhead of RC-RNN in two scenarios: 1)~\emph{multiple model}, where the system has already loaded multiple caching models, and 2)~\emph{single model}, where the system requires to load the new RNN model into the main memory. In the first scenario, when the system detects a change in the workload type~(using characterization model), the caching models are already loaded and \emph{RC-RNN} maintains its efficiency after workload change. On the other hand, the number of missed blocks in the reconfiguration process of the \emph{single model} scenario results in a small percentage of lower hit ratio in the later time periods. However, after a few periods of I/O requests, the single model RC-RNN eventually reaches the performance of the multiple model RC-RNN as shown in Figure~\ref{fig:exp_cch_dyn}.
Note that \emph{Fixed\_model} fails to accurately predict the workload access pattern.
\color{black}
Therefore, employing a model for each workload type has significant impact on the overall performance of caching architectures.

One of the drawbacks of the reconfiguration process is the memory overhead required for operating the monitoring module that analyzes the current workload every one thousand requests. In addition, in order to change the cache manager module, one needs to unload the previous and load the new RNN module into the main memory. 
\color{black}
The overhead of re-evaluating benefit values is already included in the experimental results~(Figure~\ref{fig:exp_cch_static_res} and Figure~\ref{fig:exp_cch_dyn}). Since the reconfiguration is not done frequently, its prorated overhead on the total runtime is not significant.
\color{black}

%\vspace{6mm}

\vspace{-1mm}
\section{Conclusion} \label{sec:con}
\vspace{-0.3mm}
\textcolor{black}{Current state-of-the-art SSD caching architectures} characterize running workload to identify performance-critical data pages.
Our experiments reveal that such characterization methods fail to accurately identify workload type when several applications are running simultaneously such as in virtualization environments.
Additionally, there is a significant gap between hit ratio of previous studies and maximum possible hit ratio~(Oracle model).
To reduce this gap and improve the accuracy of workload characterization, more complex methods such as RNNs need to be employed.

\color{black}
In this paper, we proposed \emph{RC-RNN}, the first reconfigurable caching architecture using RNN to characterize workloads and manage SSD cache.
\emph{RC-RNN} employs an RNN model to identify workload type.
For each workload type, a specific RNN is constructed by analyzing the decisions of \emph{Oracle} caching architecture on various workloads.
%The RNN model for currently identified workload type is dynamically loaded into \emph{RC-RNN}.
Experimental results show that \emph{RC-RNN} can identify workload type by 40\% higher accuracy compared to previous studies.
\emph{RC-RNN} also improves performance and SSD lifetime by an average of 15\% and 1.7\% (up to 50\% and 22\%), respectively.
As one of the main future work, it is possible to test the effectiveness of different alternatives in hyperparameters selection of RC-RNN models.
\color{black}
\vspace{-3mm}
\ifCLASSOPTIONcaptionsoff
\newpage
\fi



%\bibliographystyle{IEEEtran}
%\begin{thebibliography}{1}
\bibliographystyle{IEEEtran}
%\bibliography{ref}

\begin{thebibliography}{10}
\providecommand{\url}[1]{#1}
\csname url@samestyle\endcsname
\providecommand{\newblock}{\relax}
\providecommand{\bibinfo}[2]{#2}
\providecommand{\BIBentrySTDinterwordspacing}{\spaceskip=0pt\relax}
\providecommand{\BIBentryALTinterwordstretchfactor}{4}
\providecommand{\BIBentryALTinterwordspacing}{\spaceskip=\fontdimen2\font plus
\BIBentryALTinterwordstretchfactor\fontdimen3\font minus
  \fontdimen4\font\relax}
\providecommand{\BIBforeignlanguage}[2]{{%
\expandafter\ifx\csname l@#1\endcsname\relax
\typeout{** WARNING: IEEEtran.bst: No hyphenation pattern has been}%
\typeout{** loaded for the language `#1'. Using the pattern for}%
\typeout{** the default language instead.}%
\else
\language=\csname l@#1\endcsname
\fi
#2}}
\providecommand{\BIBdecl}{\relax}
\BIBdecl

\bibitem{ahmadian2}
S.~Ahmadian, F.~Taheri, M.~Lotfi, M.~Karimi, and H.~Asadi, ``Investigating
  power outage effects on reliability of solid-state drives,'' in \emph{Design,
  Automation \& Test in Europe Conference \& Exhibition (DATE)}.\hskip 1em plus
  0.5em minus 0.4em\relax IEEE, 2018, pp. 207--212.

\bibitem{kishani}
M.~Kishani, R.~Eftekhari, and H.~Asadi, ``Evaluating impact of human errors on
  the availability of data storage systems,'' in \emph{Design, Automation \&
  Test in Europe Conference \& Exhibition (DATE)}, 2017, pp. 314--317.

\bibitem{reca}
R.~Salkhordeh, S.~Ebrahimi, and H.~Asadi, ``Reca: An efficient reconfigurable
  cache architecture for storage systems with online workload
  characterization,'' \emph{IEEE Transactions on Parallel \& Distributed
  Systems (TPDS)}, no.~7, pp. 1605--1620, 2018.

\bibitem{arc}
N.~Megiddo and D.~Modha, ``Outperforming lru with an adaptive replacement cache
  algorithm,'' \emph{Journal of Computer}, vol.~37, no.~4, p. 58–65, 2004.

\bibitem{larc}
S.~Huang, Q.~Wei, J.~Chen, C.~Chen, and D.~Feng, ``Improving flash-based disk
  cache with lazy adaptive replacement,'' in \emph{Proceedings of the 29th IEEE
  Symposium on Mass Storage Systems and Technologies (MSST)}, May 2013, p.
  1–10.

\bibitem{marc}
R.~Santana, S.~Lyons, R.~Koller, R.~Rangaswami, and J.~Liu, ``To arc or not to
  arc,'' in \emph{Proceedings of the 7th USENIX Conference on Hot Topics in
  Storage and File Systems (HotStorage)}, 2015, p. 4–14.

\bibitem{azor}
Y.~Klonatos, T.~Makatos, M.~Marazakis, M.~D. Flouris, and A.~Bilas, ``Azor:
  Using two-level block selection to improve ssd-based i/o caches,'' in
  \emph{IEEE Sixth International Conference on Networking, Architecture, and
  Storage (NAS)}, 2011, pp. 309--318.

\bibitem{hystor}
F.~Chen, D.~A. Koufaty, and X.~Zhang, ``Hystor: Making the best use of solid
  state drives in high performance storage systems,'' in \emph{Proceedings of
  the International Conference on Supercomputing}.\hskip 1em plus 0.5em minus
  0.4em\relax ACM, 2011, pp. 22--32.

\bibitem{srac}
Y.~Ni, J.~Jiang, D.~Jiang, X.~Ma, J.~Xiong, and Y.~Wang, ``S-rac: Ssd friendly
  caching for data center workloads,'' in \emph{Proceedings of the 9th ACM
  International on Systems and Storage Conference}.\hskip 1em plus 0.5em minus
  0.4em\relax ACM, 2016, pp. 8:1--8:12.

\bibitem{dura}
W.-H. Kang, S.-W. Lee, B.~Moon, Y.-S. Kee, and M.~Oh, ``Durable write cache in
  flash memory ssd for relational and nosql databases,'' in \emph{Proceedings
  of the ACM SIGMOD International Conference on Management of Data}, ser.
  SIGMOD '14.\hskip 1em plus 0.5em minus 0.4em\relax ACM, 2014, pp. 529--540.

\bibitem{8550676}
R.~{Salkhordeh}, M.~{Hadizadeh}, and H.~{Asadi}, ``An efficient hybrid i/o
  caching architecture using heterogeneous ssds,'' \emph{IEEE Transactions on
  Parallel and Distributed Systems}, pp. 1--12, 2018.

\bibitem{char_tarihi1}
M.~Tarihi, H.~Asadi, A.~Haghdoost, M.~Arjomand, and H.~Sarbazi-Azad, ``A hybrid
  non-volatile cache design for solid-state drives using comprehensive i/o
  characterization,'' \emph{IEEE Transactions on Computers}, vol.~65, no.~6,
  pp. 1678--1691, June 2016.

\bibitem{belady}
L.~A. Belady, R.~A. Nelson, and G.~S. Shedler, ``An anomaly in space-time
  characteristics of certain programs running in a paging machine,''
  \emph{Commun. ACM}, vol.~12, no.~6, pp. 349--353, Jun. 1969.

\bibitem{snia}
\BIBentryALTinterwordspacing
S.~N. I.~A. (SNIA). (2017). [Online]. Available: \url{http://www.snia.org}
\BIBentrySTDinterwordspacing

\bibitem{ml4}
S.~Nishide, H.~G. Okuno, T.~Ogata, and J.~Tani, ``Handwriting prediction based
  character recognition using recurrent neural network,'' in \emph{IEEE
  International Conference on Systems, Man, and Cybernetics}, Oct 2011, pp.
  2549--2554.

\bibitem{ml2}
W.~Ali, S.~M. Shamsuddin, and A.~S. Ismail, ``Intelligent web proxy caching
  approaches based on machine learning techniques,'' \emph{Decision Support
  Systems}, vol.~53, no.~3, pp. 565 -- 579, 2012.

\bibitem{keras}
F.~Chollet, F.~Rahman, G.~de~Marmiesse, and T.~Lee, ``Keras,'' Availabe
  [Online]: https://keras.io, 2015.

\bibitem{tiering1}
R.~Salkhordeh, H.~Asadi, and S.~Ebrahimi, ``Operating system level data tiering
  using online workload characterization,'' \emph{J. Supercomput.}, vol.~71,
  no.~4, pp. 1534--1562, Apr. 2015.

\bibitem{integrating}
R.~Appuswamy, D.~C. van Moolenbroek, and A.~S. Tanenbaum, ``Integrating
  flash-based ssds into the storage stack,'' in \emph{IEEE 28th Symposium on
  Mass Storage Systems and Technologies (MSST)}, April 2012, pp. 1--12.

\bibitem{tiering2}
X.~Wu and A.~L.~N. Reddy, ``Exploiting concurrency to improve latency and
  throughput in a hybrid storage system,'' in \emph{IEEE International
  Symposium on Modeling, Analysis and Simulation of Computer and
  Telecommunication Systems}, Aug 2010, pp. 14--23.

\bibitem{Guerra}
J.~Guerra, H.~Pucha, J.~Glider, W.~Belluomini, and R.~Rangaswami, ``Cost
  effective storage using extent based dynamic tiering,'' in \emph{Proceedings
  of the 9th USENIX Conference on File and Stroage Technologies}, ser.
  FAST'11.\hskip 1em plus 0.5em minus 0.4em\relax USENIX Association, 2011, pp.
  20--20.

\bibitem{ahmadian1}
S.~Ahmadian, O.~Mutlu, and H.~Asadi, ``Eci-cache: A high-endurance and
  cost-efficient i/o caching scheme for virtualized platforms,''
  \emph{Proceedings of the ACM on Measurement and Analysis of Computing
  Systems}, vol.~2, no.~1, p.~9, 2018.

\bibitem{thrashing}
V.~Seshadri, O.~Mutlu, M.~A. Kozuch, and T.~C. Mowry, ``The evicted-address
  filter: A unified mechanism to address both cache pollution and thrashing,''
  in \emph{Proceedings of the 21st international conference on Parallel
  architectures and compilation techniques}.\hskip 1em plus 0.5em minus
  0.4em\relax ACM, 2012, pp. 355--366.

\bibitem{char_tarihi2}
M.~Tarihi, H.~Asadi, and H.~Sarbazi-Azad, ``Diskaccel: Accelerating disk-based
  experiments by representative sampling,'' \emph{SIGMETRICS Perform. Eval.
  Rev.}, vol.~43, no.~1, pp. 297--308, Jun. 2015.

\bibitem{char_disk_io}
I.~Ahmad, ``Easy and efficient disk i/o workload characterization in vmware esx
  server,'' in \emph{IEEE 10th International Symposium on Workload
  Characterization}, Sept 2007, pp. 149--158.

\bibitem{char_ms_storage}
S.~Kavalanekar, B.~Worthington, Q.~Zhang, and V.~Sharda, ``Characterization of
  storage workload traces from production windows servers,'' in \emph{IEEE 11th
  International Symposium on Workload Characterization}, Sept 2008, pp.
  119--128.

\bibitem{char_peta}
P.~Carns, R.~Latham, R.~Ross, K.~Iskra, S.~Lang, and K.~Riley, ``24/7
  characterization of petascale i/o workloads,'' in \emph{IEEE International
  Conference on Cluster Computing and Workshops}, Aug 2009, pp. 1--10.

\bibitem{char_riska}
A.~Riska and E.~Riedel, ``Disk drive level workload characterization,'' in
  \emph{USENIX Annual Technical Conference (USENIX ATC)}, 2006, pp. 97--102.

\bibitem{char_web}
M.~F. Arlitt and C.~L. Williamson, ``Web server workload characterization: The
  search for invariants,'' \emph{SIGMETRICS Perform. Eval. Rev.}, vol.~24,
  no.~1, pp. 126--137, May 1996.

\bibitem{deep}
I.~Goodfellow, Y.~Bengio, and A.~Courville, \emph{Deep Learning}.\hskip 1em
  plus 0.5em minus 0.4em\relax MIT Press, 2016,
  \url{http://www.deeplearningbook.org}.

\bibitem{ml5}
P.~Li, J.~Zhu, L.~Peng, and Y.~Guo, ``Rnn based uyghur text line recognition
  and its training strategy,'' in \emph{12th IAPR Workshop on Document Analysis
  Systems (DAS)}, April 2016, pp. 19--24.

\bibitem{label}
A.~Graves, \emph{Supervised sequence labelling with recurrent neural
  networks}.\hskip 1em plus 0.5em minus 0.4em\relax Springer, 2012, vol. 385.

\bibitem{lstm}
\BIBentryALTinterwordspacing
Z.~C. Lipton, ``A critical review of recurrent neural networks for sequence
  learning,'' \emph{CoRR}, vol. abs/1506.00019, 2015. [Online]. Available:
  \url{http://arxiv.org/abs/1506.00019}
\BIBentrySTDinterwordspacing

\bibitem{andermatt2016multi}
S.~Andermatt, S.~Pezold, and P.~Cattin, ``Multi-dimensional gated recurrent
  units for the segmentation of biomedical 3d-data,'' in \emph{Deep Learning
  and Data Labeling for Medical Applications}.\hskip 1em plus 0.5em minus
  0.4em\relax Springer, 2016, pp. 142--151.

\bibitem{cambridge}
D.~Narayanan, A.~Donnelly, and A.~Rowstron, ``Write off-loading: Practical
  power management for enterprise storage,'' \emph{Trans. Storage}, vol.~4,
  no.~3, pp. 10:1--10:23, Nov. 2008.

\bibitem{cminer}
Z.~Li, Z.~Chen, S.~M. Srinivasan, and Y.~Zhou, ``C-miner: Mining block
  correlations in storage systems.'' in \emph{FAST}, vol.~4, 2004, pp.
  173--186.

\bibitem{filebench}
A.~Wilson, ``The new and improved filebench,'' in \emph{Proceedings of the 6th
  USENIX Conference on File and Stroage Technologies}, ser. FAST.\hskip 1em
  plus 0.5em minus 0.4em\relax USENIX Association, 2008.

\bibitem{gheisari2017survey}
M.~Gheisari, G.~Wang, and M.~Z.~A. Bhuiyan, ``A survey on deep learning in big
  data,'' in \emph{2017 IEEE International Conference on Computational Science
  and Engineering (CSE) and IEEE International Conference on Embedded and
  Ubiquitous Computing (EUC)}, vol.~2.\hskip 1em plus 0.5em minus 0.4em\relax
  IEEE, 2017, pp. 173--180.

\bibitem{enhanceio}
\BIBentryALTinterwordspacing
STEC. (2017) Enhanceio ssd caching software. [Online]. Available:
  \url{https://github.com/stecinc/EnhanceIO}
\BIBentrySTDinterwordspacing

\bibitem{rmsprop}
T.~Tieleman and G.~Hinton, ``Lecture 6.5-rmsprop: Divide the gradient by a
  running average of its recent magnitude,'' \emph{COURSERA: Neural networks
  for machine learning}, vol.~4, no.~2, pp. 26--31, 2012.

\end{thebibliography}


%\vspace{-4mm}
\begin{IEEEbiography}[{\includegraphics[scale=0.36]{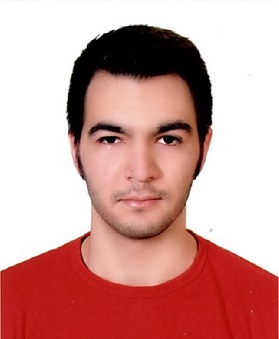}}]
{Shahriar Ebrahimi} 
received his B.Sc. degree from Sharif University of Technology, Tehran, Iran, in 2014, and the M.Sc. degree from the same university in 2016, both in computer engineering. Currently, He is a PhD candidate of computer engineering in Sharif University of Technology. His research interests include data storage systems, operating systems, and cryptographic engineering. 
\end{IEEEbiography}
%\vspace{-4mm}
\begin{IEEEbiography}[{\includegraphics[width=1in,height=1.25in,clip,keepaspectratio]{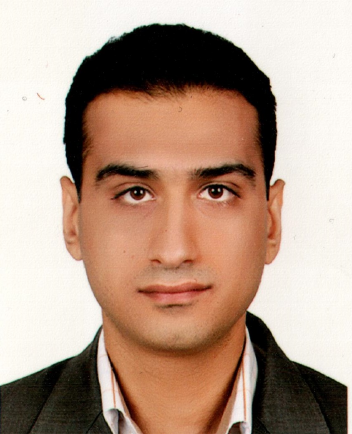}}]
{Reza Salkhordeh}
received his B.Sc. from Ferdowsi University of Mashhad in 2011, M.Sc and Ph.D. from Sharif University of Technology (SUT) in 2013 and 2018, respectively.  He was a member of PPoPP'18 artifact evaluation committee and a reviewer for Elsevier Microprocessors and Microsystems (2016) and Transactions on Computer-Aided Design of Integrated Circuits and Systems (2018). He was a member of Iran’s National Elites Foundation. His research interests are storage systems design, solid-state drives, non-volatile memories, operating systems design, and caching architectures.
\end{IEEEbiography}
\begin{IEEEbiography}[{\includegraphics[width=1in,height=1.25in,clip,keepaspectratio]{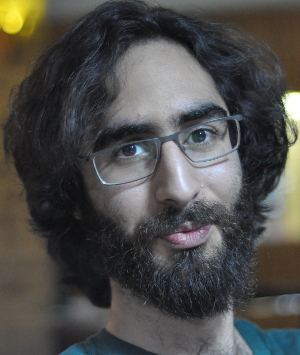}}]{Seyed Ali Osia}
received his B.Sc. degree in Software Engineering from Sharif University of Technology in 2014. He is currently a Ph.D. candidate at the department of computer engineering, Sharif University of Technology. His research interests includes Statistical Machine Learning, Deep Learning, Privacy and Computer Vision.
\end{IEEEbiography}
\begin{IEEEbiography}[{\includegraphics[width=1in,height=1.25in,clip,keepaspectratio]{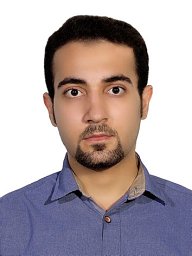}}]{Ali Taheri}
received his B.Sc. degree in Software Engineering from Shahid Beheshti University in 2015. He received his M.Sc. degree in Artificial Intelligence from Sharif University of Technology in 2017. His research interests includes Deep Learning and Privacy.
\end{IEEEbiography}
%\vskip -3.5\baselineskip plus -1fil
%\vfill
%\vspace{-.6cm}
\begin{IEEEbiography}[{\includegraphics[width=1in,height=1.25in,clip,keepaspectratio]{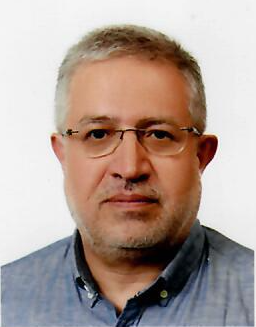}}]
{Hamid R. Rabiee} (SM’07) received his Ph.D. in Electrical and Computer Engineering from Purdue University, West Lafayette, IN, in 1996. From 1993 to 1996 he was a Member of Technical Staff at AT\&T Bell Laboratories. From 1996 to 1999 he worked as a Senior Software Engineer at Intel Corporation. He was also with PSU, OGI and OSU universities as an adjunct professor of Electrical and Computer Engineering from 1996-2000.
Since September 2000, he has joined Sharif University of Technology, Tehran, Iran. He was also a visiting professor at the Imperial College of London for the 2017-2019 academic years. His research interests include statistical machine learning, Bayesian statistics, data analytics and complex networks with applications in social networks, multimedia systems, cloud and IoT privacy, bioinformatics, and brain networks.
\end{IEEEbiography}
\begin{IEEEbiography}[{\includegraphics[width=1in,height=1.25in,clip,keepaspectratio]{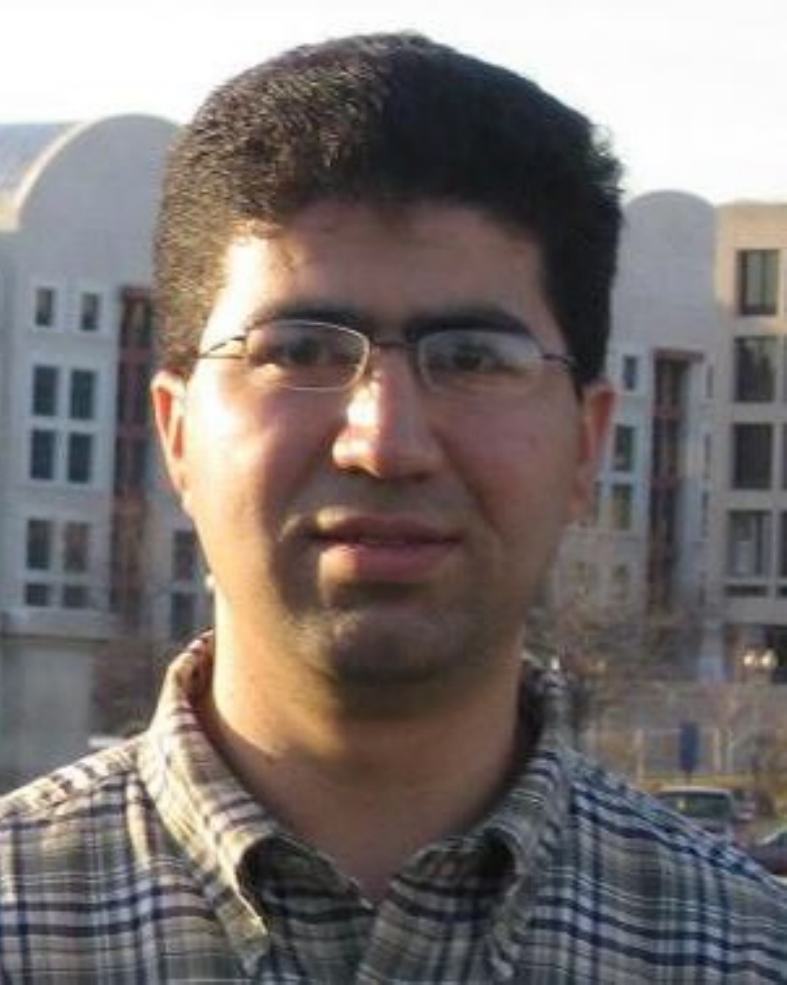}}]
{Hossein Asadi}
(M'08, SM'14) received the PhD degree in electrical and computer engineering from Northeastern University, Boston, MA, USA, in 2007. 
He was with EMC Corporation, Hopkinton, MA, as a research scientist and senior hardware engineer, from 2006 to 2009. He has been with the Department of Computer Engineering, SUT, since 2009, where he is currently a full professor. His current research interests include data storage systems and networks, solid-state drives, operating system support for I/O and high-performance computing.
More recently, he received the Best Paper Award at IEEE/ACM Design, Automation, and Test in Europe (DATE) in 2019. He has served as a guest editor of IEEE Transactions on Computers, an Associate Editor of Microelectronics Reliability, a Program Co-Chair of CADS2015, and the Program Chair of CSI National Computer Conference (CSICC2017). He is a senior member of the IEEE. 
\end{IEEEbiography}


%\vfill



\end{document}